\documentclass[10pt,aps,prb,twocolumn,superscriptaddress]{revtex4-2}
\usepackage{graphicx}
\usepackage[colorlinks, linkcolor=blue]{hyperref}
\hypersetup{colorlinks,allcolors=blue}
\usepackage{physics}
\usepackage{mathrsfs}
\usepackage{bm}
\usepackage{amssymb}
\usepackage[normalem]{ulem}
\usepackage{breqn}
\begingroup\makeatletter
\catcode`,=\active
\global\let\breqn@comma,
\protected\gdef,{\ifmmode\expandafter\breqn@comma\else\expandafter\active@comma\fi}
\endgroup

\usepackage[capitalise]{cleveref}

\newcommand{\bcen}{\begin{center}}
\newcommand{\ecen}{\end{center}}
\newcommand{\btab}{\begin{tabular}}
\newcommand{\etab}{\end{tabular}}
\newcommand{\bdes}{\begin{description}}
\newcommand{\edes}{\end{description}}

\newcommand{\beq}{\begin{equation}}
\newcommand{\eeq}{\end{equation}}
\newcommand{\bea}{\begin{eqnarray}}
\newcommand{\eea}{\end{eqnarray}}

\newcommand{\bary}{\begin{array}}
\newcommand{\eary}{\end{array}}
\newcommand{\benum}{\begin{enumerate}}
\newcommand{\eenum}{\end{enumerate}}
\newcommand{\bitem}{\begin{itemize}}
\newcommand{\eitem}{\end{itemize}}

\newcommand{\al}{\alpha}

\newcommand{\eqn}[1] {eqn.~(\ref{#1})}

\newcommand{\sect}[1] {Section~\ref{#1}}

\newcommand{\Fig}[1]{Fig.~\ref{#1}}

\makeatletter

\newcommand{\Rmnum}[1]{\expandafter\@slowromancap\romannumeral #1@}
\makeatother

\begin{document}

\title{Emergent topological phase from a one-dimensional network of defects}

\author{Rahul Singh}
\email{rahulsingh21@iitk.ac.in}
\affiliation{Department of Physics, Indian Institute of Technology Kanpur, Kalyanpur, Uttar Pradesh 208016, India}

\author{Ritajit Kundu}
\email{ritajitk@imsc.res.in}
\affiliation{Department of Physics, Indian Institute of Technology Kanpur, Kalyanpur, Uttar Pradesh 208016, India}
\affiliation{Institute of Mathematical Sciences, CIT Campus, Chennai 600113, India}

\author{Arijit Kundu}
\email{kundua@iitk.ac.in}
\affiliation{Department of Physics, Indian Institute of Technology Kanpur, Kalyanpur, Uttar Pradesh 208016, India}

\author{Adhip Agarwala}
\email{adhip@iitk.ac.in}
\affiliation{Department of Physics, Indian Institute of Technology Kanpur, Kalyanpur, Uttar Pradesh 208016, India}

\begin{abstract}
Symmetry-protected topological phases of matter, characterized by non-trivial band topology, are spectrally gapped and show non-trivial boundary phenomena.
Here, we show that scattering states when interjected by an array of periodically modulated defects can result in emergent topological 
phases whose properties can be tuned by modulating the defect strengths. We dub this the Su-Schrieffer-Heeger network. 
We show that a scattering-matrix network model can capture the emergent symmetries and nontrivial winding of the quasienergy bands, which lead to distinct transport signatures and can be further periodically driven to realize a robust Thouless charge pump.
We show that a microscopic lattice model embedded with a defect superlattice yields Bloch minibands that directly map to the network problem.  
We further verify that the physics we report is stable to disorder and point out concrete experimental solid-state platforms where it is readily realizable. 
Our work, in contrast to engineering atomic Hamiltonians, shows that defect engineering on metallic platforms can lead to emergent topological phases of quantum matter. 
\end{abstract}

\maketitle

\section{Introduction}

Symmetry-protected topological phases of matter have been at the forefront of condensed matter research, driven both by fundamental interest and by the promise of fault-tolerant quantum technologies.
While initially developed within the framework of band theory for crystalline solids, topological phases have since been identified in a remarkable range of settings, including amorphous systems~\cite{Agarwala_PRL_2017, Xu2017NatComm, Mitchell2018NatPhys, Costa2019NanoLett, Li2009PRL}, quasicrystals~\cite{Kraus2012PRL, KrausZilberberg2012PRL, Fulga2016PRL}, tree structures~\cite{ManojShenoy2023}, and scattering networks~\cite{chalker1988percolation,AltlandZirnbauer2015PRL,Albert2015PRL,ChalkerPRL2020Floquet}.
However, a recurring challenge, especially in one dimension, where Majorana-based proposals for topological quantum computation have attracted intense effort~\cite{Kitaev_PU_2001, Lutchyn2010, Oreg2010, Mourik2012, Chang2015}, is that microscopic disorder and decoherence can obscure or destroy the desired topological signatures.
This motivates the search for alternative strategies that exploit, rather than suffer from, the presence of impurities.

A natural strategy is \emph{defect engineering}: deliberately introducing a controlled arrangement of impurities or potentials to sculpt the low-energy physics of a host system~\cite{Smith_2019,Reshodko_2019,Smith_2021,De_Martino_2023,Miranda_2024,Brey_2025,vzlabys2025emergent}.
This philosophy underlies the realization of the Su-Schrieffer-Heeger (SSH) model~\cite{Su_PRL_1979} in quantum-dot arrays~\cite{Meier_2016, Slot2017, Kiczynski_2022} and, more recently, the engineering of one-dimensional topological phases via long-wavelength Moir\'{e} potentials in carbon nanotubes~\cite{Zhou_2024, chen2025tunable, Yu2023PRL, Zhao2020PRL}.
In a complementary vein, scattering between low-energy modes in metallic systems has a long history of providing deep insights into the physics of the quantum Hall effect~\cite{Lee_PRL_1993, Chalker1996networkmodels, Cho_PRB_1997, kramer2005random, Snyman_PRB_2008, GruzbergPRLFinite-Size, GruzbergPRBkagome}, synthetic quantum systems~\cite{PhysRevX2015Measurement, Chalker2001Thermalnetwork, Delplace_sciadv_2023, Delplace_nature2021}, and twisted-graphene platforms~\cite{Pasek2014Networkmodels, Huang_PRL_2018, yoo2019atomic, BeuleRPL2020, Bilayergraphenenetwork2021, VakhtelPRB2022, Kundu_PRB_2024}.
Yet the potential of coherent impurity scattering to generate topological phases in an otherwise featureless metallic wire has remained largely unexplored.

In this work, we show that a periodic superlattice of impurities deposited on a one-dimensional metallic wire gives rise to emergent topological phases whose properties are governed entirely by the impurity strengths. Crucially, the emergent symmetries of the low-energy theory differ from the microscopic symmetries of the host metal: while the bare Hamiltonian belongs to symmetry class  \cite{Altland_PRB_1997, Kitaev_AIP_2009}, which is topologically trivial in one dimension, the impurity superlattice induces an effective sublattice symmetry, promoting the system to class BDI and enabling nontrivial winding numbers. We formulate the problem within a scattering-network framework in which Fermi-surface electrons coherently scatter off the impurity array, resulting in a minimal model similar to SSH physics. The network hosts topological and trivial phases separated by a gap-closing transition tuned by the ratio of impurity strengths, and it supports localized edge excitations, quantized Thouless charge pumping, and spectral stability against disorder.

To corroborate these findings, we construct a microscopic tight-binding lattice model with a defect superlattice and show that its Bloch minibands map directly onto the quasienergy spectrum of the SSH network.
The mapping provides a concrete dictionary between the impurity parameters of the lattice model and the scattering-matrix parameters of the network description, establishing that the two viewpoints are quantitatively equivalent.
We further identify solid-state platforms, such as arrays of quantum point contacts in integer quantum Hall systems, where the SSH network can be realized directly.

The remainder of the paper is organized as follows. In \sect{sec:defnetwork}, we introduce the impurity model and present transport calculations that reveal the formation of transmission bands and gaps. In \sect{sec:topological_network}, we develop the scattering-matrix formalism, establish the topological character of the SSH network through winding-number calculations, edge-mode analysis, and a Thouless charge-pump protocol. In \sect{sec:latticereal}, we introduce the tight-binding lattice realization and its Wannierization, and in \sect{sec:mapping} we establish the microscopic-to-network mapping. In \sect{sec:realisation}, we discuss candidate experimental platforms. Finally, in \sect{sec:outlook} we summarize our findings and outline future directions.

\begin{figure}
    \centering
    \includegraphics[width=1\linewidth]{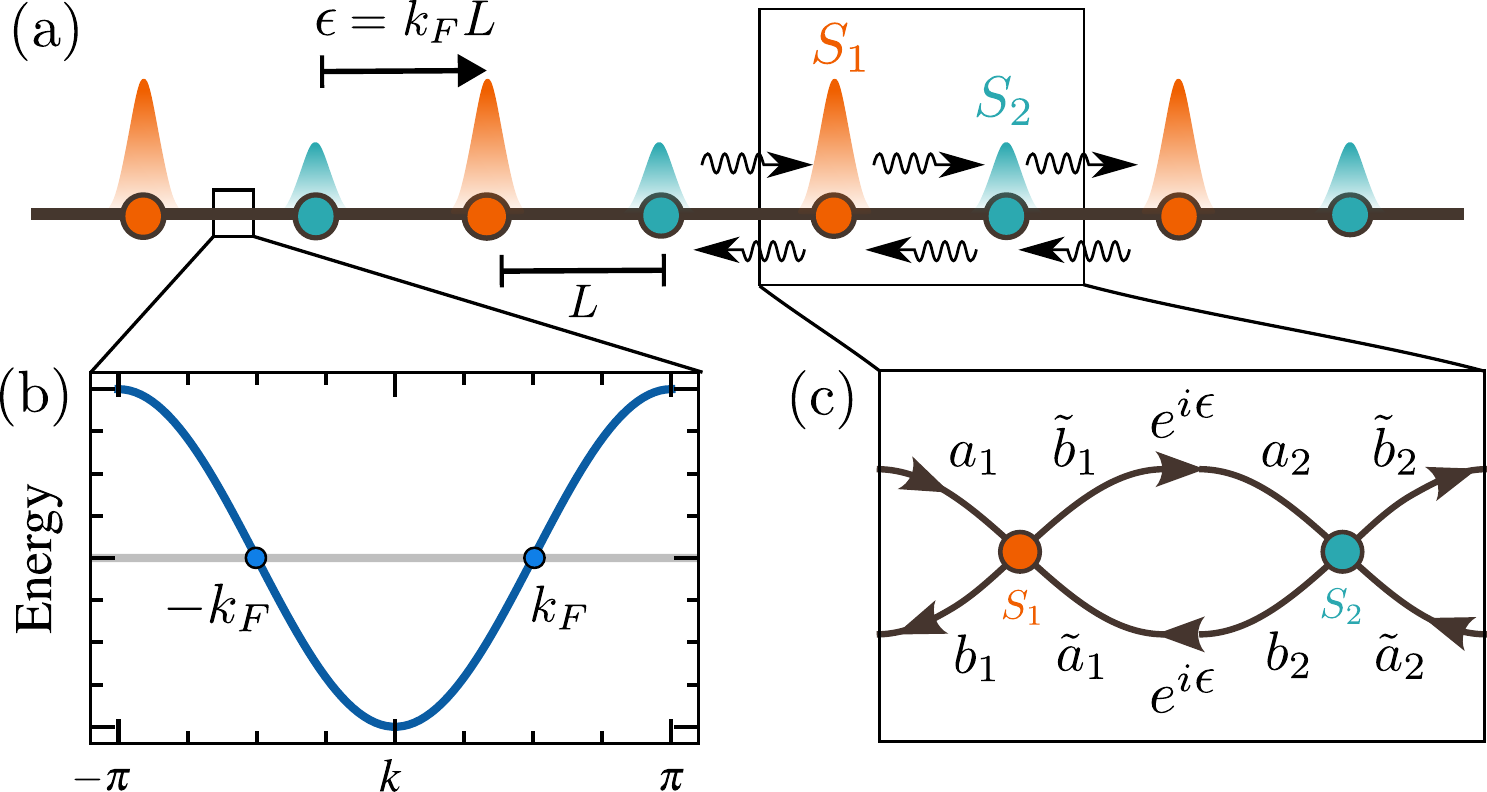}
    \caption{\textbf{Network model}. (a) Schematic of a metallic chain doped with spatially modulated defects, of alternating potential strengths $V_1, V_2$ with equal distance $L$ between them. A scattering state characterized by momentum $k_{\rm{F}}$ acquires a dynamical phase $\epsilon=k_{\rm{F}}L$ during free propagation between successive scatterers. (b) A schematic of the energy dispersion of a metallic chain, the two points $\pm k_{\rm{F}}$ correspond to the scattering waves traveling in opposite directions. (c) A unit cell with two successive scattering matrices $S_1$ and $S_2$, with their corresponding incoming and outgoing wave function amplitudes and the dynamical phase $\epsilon$ accumulated between them.
}
    \label{fig:Fig_1}
\end{figure}

\section{Impurity Model}
\label{sec:defnetwork}

We consider a one-dimensional network that consists of periodically placed impurities on a metallic wire with an inter-impurity separation $L$ [see \Fig{fig:Fig_1}(a)]. This natural length scale can be related to a dynamical phase of $\epsilon = k_F L$, which a fermion carries as it scatters from one impurity to another [see \Fig{fig:Fig_1}(b)]. Here, the $k_F$ is the Fermi momentum determined by the Fermi energy $E_F$ of the metal. The scatterers can be characterized by a scattering matrix $S_\alpha$, which is generically dependent on both the incident energy and the impurity strengths $V_\alpha$.  A periodic arrangement of two different impurity strengths $V_1$ and $V_2$ leads to two different scattering matrices $S_1$ and $S_2$, respectively [see \Fig{fig:Fig_1}(c)] represented by the following unitary matrix:
\begin{equation}
S_{\alpha} =\begin{pmatrix}
r_{\alpha} &t'_{\alpha} \\
t_{\alpha}&r'_{\alpha}
\end{pmatrix}
\label{S_mat_form}
\end{equation}
where $r_{\alpha},\, r'_{\alpha},\, t_{\alpha},\, t'_{\alpha}$ are complex numbers with $|r_{\alpha}|^2+|t_{\alpha}|^2= |r'_{\alpha}|^2+|t'_{\alpha}|^2=1$. 
The scattering matrix relates the incoming and outgoing wavefunction amplitudes [see \Fig{fig:Fig_1}(c)]
\begin{equation}
\begin{pmatrix}
b_{\alpha}\\
\tilde{b}_{\alpha}    
\end{pmatrix}
=S_{\alpha}
\begin{pmatrix}
a_{\alpha}\\
\tilde{a}_{\alpha}
\end{pmatrix}.
\label{S_mat_def}
\end{equation}
For scalar impurities with both time-reversal and inversion symmetry, such that $r_\alpha = r'_\alpha$ and $t_\alpha = t'_\alpha$ (see \cref{scatsymm}), we evaluate the transmission ($T$) through such a network. We numerically compute the transmission $T$ through the network using Chebyshev’s identity, which is particularly useful for calculating transmission through an array of periodically placed scatterers \cite{markos2008wave} (see \cref{transmission} for details).

\begin{figure}
    \centering
    \includegraphics[width=1\linewidth]{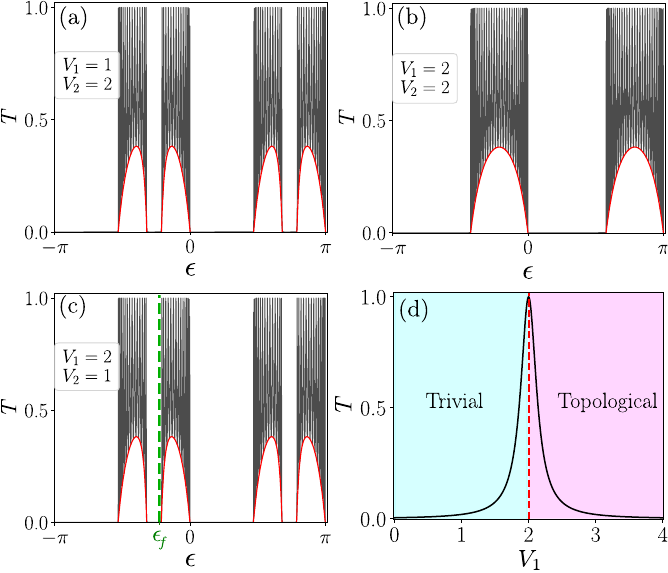}
    \caption{\textbf{Transmission through a defect network}.  Transmission $T$ as a function of the dynamical phase $\epsilon$ for a network consisting of $15$ unit cells with (a) $V_1=1,V_2=2$. (b) $V_1=2,V_2=2$. (c) $V_1=2,V_2=1$, identical to panel~(a). (d) Transmission $T$ as a function of the scatterer strength $V_1$ for fixed $V_2=2$ and fixed dynamical phase $\epsilon_{f}$ shown by vertical dashed line in panel~(c). 
    }     \label{fig:Fig2}
\end{figure}

Our numerical results are shown in \Fig{fig:Fig2}(a--c), where for different combinations of $(V_1, V_2)$ the transmission is shown as a function of $\epsilon = (k_F L) \bmod{2\pi}$. For $V_1 \neq V_2 $, there are four transmission bands separated by finite gaps [panels~(a) and (c)]. In contrast, for $V_1=V_2$, adjacent transmission bands merge, resulting in only two transmission bands [panel~(b)]. We next examine the dependence of transmission on the impurity strength $V_2$ for fixed $V_1$ and fix 
$\epsilon_f \equiv \epsilon= -0.66$ to a band-edge, as indicated by the vertical dashed line in panel~(c). The transmission is maximized when the impurities are identical, $V_1=V_2$, and is progressively suppressed as the asymmetry between the impurity strengths increases, as shown in \Fig{fig:Fig2}(d). Thus, tuning the impurity strengths provides an efficient control knob on the transport properties of the scattering network.
As we discuss below, the regimes $V_1<V_2$ and $V_1>V_2$ are topological and trivial, respectively.

In the next section, we discuss the formalism to calculate these transport quantities, to discuss the origin of such bands and their gaps, and further discuss
the topological character of such networks. In the process, we introduce a minimal model of a non-trivial network: a Su-Schrieffer-Heeger (SSH) equivalent of a topological network.

\begin{figure*}
    \centering
    \includegraphics[width=0.9\linewidth]{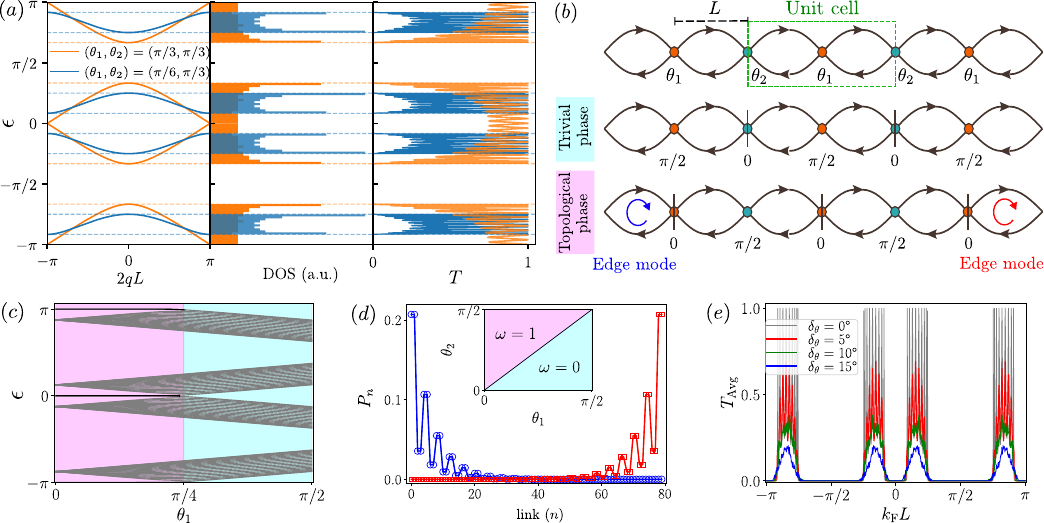}
    \caption{
\textbf{Topological phases in SSH network}. (a) Left: The plot of the quasi energy bands $\epsilon$ as a function of momentum $q$ for two representative values of $(\theta_1,\theta_2)$. Middle: Density of states as a function of $\epsilon$. Right: Transmission $T$ through an open network as a function of dynamical phase $\epsilon$, with $N=10$ unit cells and $\phi_1=\phi_2=0$ (see \eqn{General_s_mat}). (b) Schematics of an open network with alternate scattering nodes separated by distance $L$ with red and blue circles with parameters $\theta_1$ and $\theta_2$, respectively. The green box indicates the unit cell of the network. The bottom two figures show the ``atomic limits" of of the network trivial ($\theta_1= \pi/2, \theta_2 = 0$) and topological phases ($\theta_1= 0, \theta_2 = \pi/2$). The vertical bars separate the connected loops by complete reflection. (c) Quasi energy $\epsilon$ as a function of $\theta_1$ with open boundary condition of network, for $\theta_2= \pi/4$ and $N=20$ unit cells. When $\theta_1<\theta_2$, the spectrum have degenerate quasi energies at $\epsilon=0$ and $\epsilon= \pi $. (d) Plot of the probability density $P_n$ as a function of link index $n$ corresponding to $\epsilon=0$ with $\theta_1= \pi/4, \theta_2=\pi/3 $ and $N=20$. Inset figure: Winding number $w$ as a function of $\theta_1$ and $\theta_2$. $(e)$ Plot of average transmission $(T_{\rm{Avg}})$ as a function of dynamical phase $k_{\rm{F}}L$ in a disordered system with $\theta_1= 30^{\circ} + \Delta_{\theta_1}$ and $ \theta_2= 60^{\circ} + \Delta_{\theta_2}$ where $\Delta_{\theta_1}, \Delta_{\theta_2}$ are uniformly chosen from $[-\delta_{\theta}, \delta_{\theta} ]$. Configuration averaging is performed over $100$ realizations with $N=10$ unit cells in the network. 
    } 
    \label{fig:Fig3}
\end{figure*}

\section{SSH Network }
\label{sec:topological_network}
    
\subsection{Quasienergy band structure}
\label{subsec:Quasibands}
We discuss the details of the general scattering-matrix formalism \cite{Delplace_nature2021, Pasek2014Networkmodels, Kundu_PRB_2024} as applied to this one-dimensional system, characterized by a periodic sequence of impurities with strengths $V_1$ and $V_2$. $S_\alpha$ [see \cref{S_mat_form}] can be fully characterized by two parameters $\theta_\alpha$ and $\phi_\alpha$:
\begin{align}
	S_{\al} =
	 e^{i \phi_{\al}}\begin{pmatrix}   \cos \theta_{\al}   & i \sin \theta_{\al} \\  i \sin \theta_{\al} & \cos \theta_{\al} \end{pmatrix},  \label{General_s_mat}
\end{align}
where $\theta_\alpha \in [0,\pi/2]$ and $\phi_\alpha \in [0,2\pi)$ where $\alpha\in\{1,2\}$ labels the two types of scatterers in the network. Here $\theta_\alpha$ parametrizes the scattering strength via $|r_\alpha|=\cos\theta_\alpha$ and $|t_\alpha|=\sin\theta_\alpha$. The limits $\theta_\alpha= 0$ correspond to perfect reflection, whereas $\theta_\alpha=\pi/2$ describes a transparent scatterer. The scattering matrix of the $j$-th unit-cell of the network [as shown in \cref{fig:Fig_1}(c)], is given by the direct sum ($\oplus$) of the two scattering matrices:
\begin{equation}
    S^{(j)}_{\rm{unit}}=  
   S^{(j)}_1 \oplus 
   S^{(j)}_2
    \label{S_unit_tensor}
\end{equation}
During the scattering event, $S^{(j)}_{\rm{unit}}$ acts on the incoming amplitudes ${\bm a}^{(j)}= (a^{(j)}_1, \tilde{a}^{(j)}_1,a^{(j)}_2, \tilde{a}^{(j)}_2)^{\rm{T}}$ and relates with the outgoing amplitudes ${\bm b}^{(j)}= (b^{(j)}_1, \tilde{b}^{(j)}_1,b^{(j)}_2, \tilde{b}^{(j)}_2)^{\rm{T}}$, i.e.,
\begin{equation}
    \bm{b}^{(j)}= S^{(j)}_{\rm{unit}} \bm{a}^{(j)},
    \label{S_unit_relaion}
\end{equation}
here $(\cdot)^{\mathrm{T}}$ represents transpose of the matrix. After scattering, electrons propagate freely over a distance $L$ along the network before the next scattering event. These scatterer-free segments are referred to as the \emph{links} of the network. During propagation along a link, an electron acquires a dynamical phase $e^{i\epsilon}$, with $\epsilon \in (-\pi,\pi]$. Thus, in a physical system, $\epsilon = (k_F L) \bmod{2\pi}$, as was introduced in the last section.  As a result, the incoming amplitudes of the $j$-th unit cell are related to the outgoing amplitudes of neighboring cells as
\begin{align}
    (a^{(j)}_{1}, &\tilde{a}^{(j)}_{1}, a^{(j)}_{2}, \tilde{a}^{(j)}_{2})^{\rm{T}} =e^{i\epsilon} (\tilde{b}^{(j-1)}_{2}, b^{(j)}_{2}, \tilde{b}^{(j)}_{1}, {b}^{(j+1)}_{1})^{\rm{T}}
    \label{eq:links}
\end{align}

For a translationally invariant system, Bloch’s theorem implies
$
b^{(j\pm1)}_{\alpha} = {b}^{(j)}_{\alpha} e^{\pm 2iqL}
$, where the unit cell has length $2L$ and $q$ is the Bloch-momentum. Using \cref{S_unit_relaion,eq:links} and suppressing the unit-cell superscript $(j)$, we obtain the eigenvalue equation
\begin{align}
     &S_{\mathrm{unit}} M \bm b = e^{-i \epsilon} \bm  b, \label{eq:eigen}\\
     &M = 
     \begin{pmatrix}
         0 & 0 & 0 & e^{-2 i q L} \\
         0 & 0 & 1 & 0 \\
         0 & 1 & 0 & 0 \\
         e^{2 i q L } & 0 & 0 & 0
     \end{pmatrix}
\end{align}
By solving the eigen-problem in \cref{eq:eigen}, we obtain the quasi energy bands of the dynamical phase $\epsilon(q)$. 
In the left panel of \cref{fig:Fig3}(a), we show numerical plots of the quasi energy bands $\epsilon$ for representative parameters. Since the overall phases $\phi_{1,2}$ only shift the spectrum, we set $\phi_1=\phi_2=0$ without loss of generality; results for finite $\phi_{1,2}$ are presented in the \cref{sec:finiteshift}.

Generally, for $\theta_1\neq \theta_2$ there exist four quasi energy bands, of which two are independent due to a twofold \emph{phase-rotation symmetry} \cite{Delplace_2017}. When $\theta_1 \neq \theta_2$, finite gaps open at $\epsilon=0$ and $\epsilon=\pi$. This occurs because identical scatterers imply a unit cell that is twice the size of the primitive cell, resulting in an artificially folded Brillouin zone. Interestingly, the gaps at $\epsilon=\pm\pi/2$ never close for any finite scatterer strength. The density of states (DOS) of the quasi energy bands is shown in the middle panel of \Fig{fig:Fig3}(a). Using the transfer matrix formalism (as discussed in Appendix~\ref{transmission}), we compute the transmission $T$, as shown in the right panel of \Fig{fig:Fig3}(a). Transmission occurs only when the dynamical phase $\epsilon$ lies within the quasi energy bands obtained from \cref{eq:eigen}. If $\epsilon$ belongs to the band gaps, the transmission is exponentially suppressed. This shows that for a fixed $\epsilon = (k_F L) \bmod{2\pi}$, given the impurity strengths, one can tune the system to be either in a conducting or an insulating state as was shown in the \Fig{fig:Fig2}.

\subsection{Edge excitations}

We now consider the network under open boundary conditions (OBC) (see top panel in \Fig{fig:Fig3}(b)). While it is straightforward to implement OBC in Hamiltonian systems, in a network, the reflection at the edge introduces an ambiguity in the reflection phase. The exact boundary condition depends on the system's truncation and must be determined from the microscopic model. Here, we set the reflection phase at the edge to zero, with a unit amplitude (infinite boundary potential). 

We first examine the extreme limits of the network in parameter space $(\theta_1, \theta_2)$, which we refer to as ``atomic limits" of the network. In one limit, the $\alpha=1$ scatterers are fully transparent and the $\alpha=2$ scatterers are fully reflecting, corresponding to $\theta_1 = \pi/2$ and $\theta_2 = 0$ (see~\Fig{fig:Fig3}(b) middle panel). In this case, the network decomposes into independent partitions, each containing the two scatterers and four links effectively and separated by vertical bars.  In the other limit, the $\alpha=1$ scatterers are fully reflecting and the $\alpha=2$ scatterers are fully transparent, corresponding to $\theta_1 = 0$ and $\theta_2 = \pi/2$ (see \cref{fig:Fig3}(b) bottom panel). Here, the bulk still partitions into two scatterers and four links, but at the edges, the network decomposes to a single scatterer with two links. This is reminiscent of the Su-Schrieffer-Heeger physics \cite{Su_PRB_1980}, where the Hamiltonian splits up in a similar way. These edge partitions correspond to \emph{edge modes} of the network, and as we will show below. 

We obtain the quasi energy spectrum following an approach similar to \cref{eq:eigen}, adapted to the open geometry shown in \cref{fig:Fig3}(b) (see Appendix~\ref{subsec:Appen_S_eigen_value_open} for details). Numerical results for representative parameters are shown in \cref{fig:Fig3}(c). When $\theta_1 < \theta_2$, twofold degenerate subgap modes appear at phases $\epsilon=0$ and $\epsilon=\pi$. These modes are localized at the edges of the network, as shown in \cref{fig:Fig3}(d), and decay exponentially into the bulk (see \cref{app:scateff}). Close to the gap-closing point $ (\theta_1 \approx \theta_2)$, the localization length $(\xi)$ diverges as $\xi \sim 1/|\theta_1-\theta_2|$. The probability density on $n$-th link of the network is given by $P_n = \bm b^\dagger \mathcal I_n \bm b$,
where $\bm b$ is the eigenvector of the subgap state and $\mathcal I_n$ projects onto the $n$-th link of the network.
The resulting localization profiles of edge states are shown in \cref{fig:Fig3}(d). In contrast, when $\theta_1=\theta_2$ the gaps close, and for $\theta_1>\theta_2$ no subgap modes are present, as seen in \cref{fig:Fig3}(c).

\subsection{Winding number}

The appearance of edge modes suggests a topological origin. Such modes are a hallmark of topological insulators, where the bulk is insulating while localized subgap states appear at the boundaries. To make this connection explicit, we compute the winding numbers of the bulk bands under periodic boundary conditions (PBC). We use the eigenvectors obtained from \cref{eq:eigen} and discretize the Brillouin zone into 
$N$ points with momenta $2 q_i L = 2\pi i / N$, for $i=0,\ldots,N-1$, with spacing $2\Delta q L = 2\pi / N$. The winding number is given by
\begin{align}
w = -\frac{1}{\pi} \, \mathrm{Im}
\left[
\log \prod_{i=0}^{N-1}
\bm b^\dagger(q_i)\, \bm b(q_i+\Delta q)
\right].
\end{align}

As shown numerically in the inset of \cref{fig:Fig3}(d), the winding number equals $w=1$ for $\theta_1 < \theta_2$, coinciding with the presence of localized edge states. In contrast, for $\theta_1 > \theta_2$ the winding number vanishes ($w=0$) and no localized edge states are present, corresponding to a trivial phase. At $\theta_1=\theta_2$, the quasi energy band gap closes, and the winding number becomes ill-defined. This demonstrates that tuning the scattering strengths drives a transition between topological and trivial phases. In this sense, this network mirrors the physics of the SSH model as usually understood in a Hamiltonian system.  

\subsection{Stability to disorder}

To examine the stability of this topological phase, we next investigate the effect of disorder in the scatterer strengths on the network's transmission $(T)$. We model the disorder by choosing $\theta_{\alpha}$ from a uniform distribution with mean $\bar\theta_{\alpha}$ and width $2\delta_{\theta}$, about the mean. With this, we compute the two-terminal scattering matrix of the full network for each \emph{disorder configuration} (see \cref{sec:Appen_Rediffer_prod}).
 A numerical plot of disorder configuration averaged transmission $T_{\rm{Avg}}$ as a function of the dynamical phase $\epsilon= k_{\rm{F}}L$, for different disorder-strengths $(\delta_{\theta})$ is shown in \Fig{fig:Fig3}(e). We find that, with increasing disorder strength, the average transmission $(T_{\rm{Avg}})$ is progressively suppressed. Interestingly, the transmission band gaps remain intact, indicating robustness of the topological quasi-energy bands against small disorder. 

\begin{figure}
\centering
\includegraphics[width=1\linewidth]{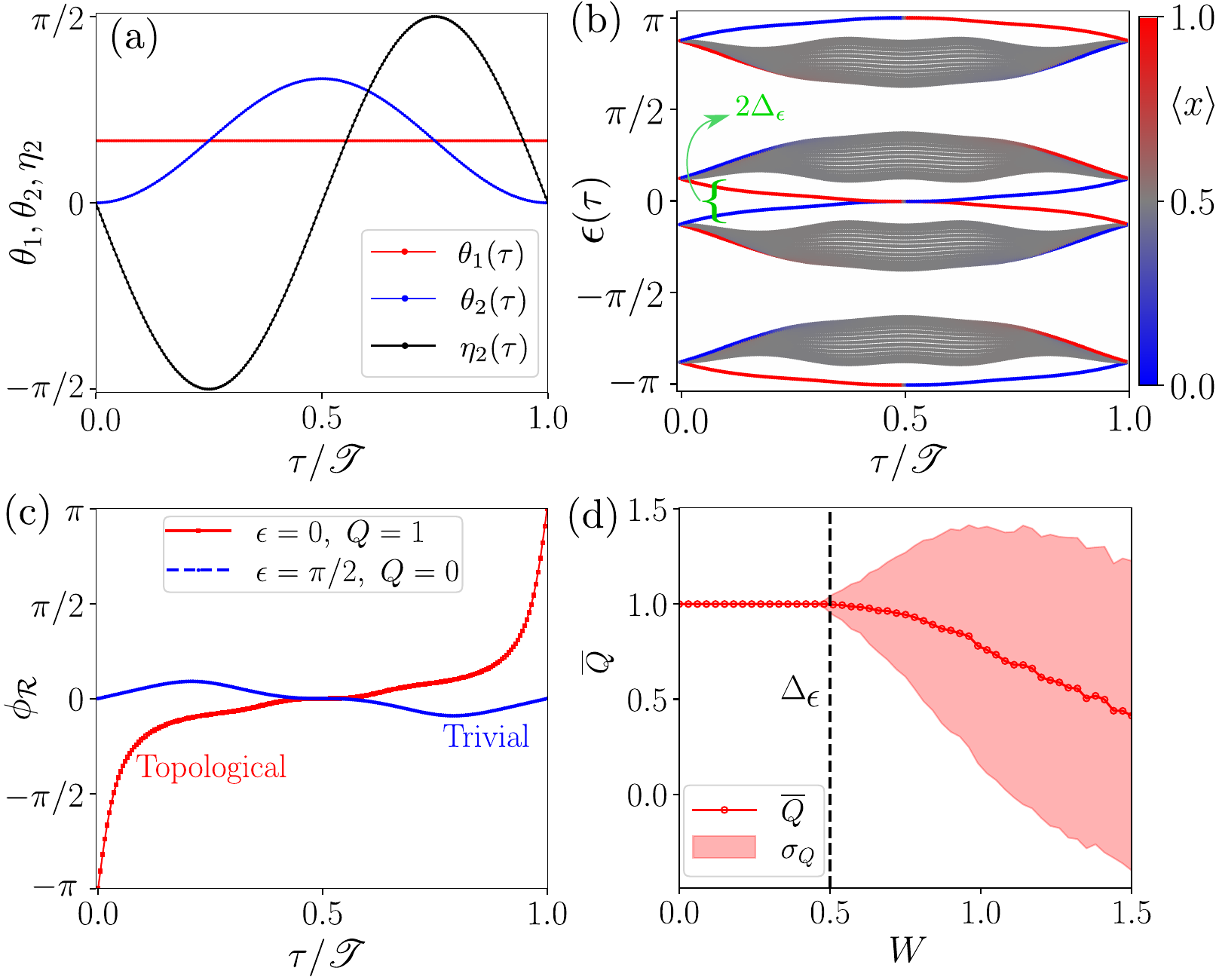}
\caption{
\textbf{Charge pumping in the network model.} 
(a) Driving protocol (\cref{eqn:driving_protocol}) used for charge pumping. 
(b) Instantaneous quasienergy spectrum \(\epsilon(\tau)\) with open boundary conditions, colored by the expectation value \(\langle x \rangle\), showing spectral flow across the gap at \(\epsilon=0\) (\(\theta_0=\pi/4\), \(\eta_0=\pi/2\), \(\mathcal{N}=15\)). 
(c) Instantaneous phase \(\phi_{\mathcal{R}}\): for \(\epsilon=0\) (red) the phase winds once, giving \(Q=1\), while for \(\epsilon=\pi/2\) (blue) it does not wind, giving \(Q=0\). 
(d) Disorder-averaged pumped charge \(\overline{Q}\) as a function of disorder strength \(W\), averaged over \(3\times10^4\) disorder realizations with random \(\epsilon\) on each link. The red shaded region indicates the standard deviation. The vertical dashed line marks \(W \approx \Delta_{\epsilon}\), up to which \(\overline{Q}=1\); here \(2\Delta_{\epsilon}\) is the bulk gap around \(\epsilon=0\).
}
    \label{fig:Fig4}
\end{figure}

\subsection{Thouless pump}
\label{sec:Thouless_pump}

A tell-tale sign of a topological band in one-dimension is the realization of the Thouless pump \cite{Thouless_pumping_83}, where an adiabatic route in the parameter space leads to the movement of a unit charge from one end of the system to the other end. The route is chosen in a way that the system remains gapped and the instantaneous band structure shows an edge crossing \cite{asboth2016short}. To further reveal the topological character of the quasi-energy bands in the SSH network, we now implement an adiabatic protocol for the scattering network. 

We introduce a new parameter, $\eta_{\alpha}$, in the scattering matrix, which arises from the lack of inversion symmetry, akin to the mass parameter in the Rice-Mele model \cite{RiceMele_PRL_1982}. Thus modified scattering matrix is:
\begin{align}
	\tilde{{S}}_{\alpha} =
	e^{i \phi_{\alpha}}
	\begin{pmatrix}
		e^{-i\eta_{\alpha}} \cos \theta_{\alpha} & i \sin \theta_{\alpha} \\
		i \sin \theta_{\alpha} & e^{i\eta_{\alpha}} \cos \theta_{\alpha}
	\end{pmatrix},
	\label{Sym_broken_S_mat}
\end{align}

We now describe the adiabatic driving protocol. During the drive, only the parameters of $\tilde{{S}}_{2}$ are varied, while those of $\tilde{{S}}_{1}$ are kept fixed:
\begin{align}
&\theta_1 = \theta_0, \qquad \eta_1 = 0, \qquad \phi_1=\phi_2=0\nonumber \\
&\theta_2 = \theta_0 \bigl[1 - \cos(2\pi \tau / \mathscr{T})\bigr], \nonumber \\
&\eta_2 = -\eta_0 \sin(2\pi \tau / \mathscr{T}),
\label{eqn:driving_protocol}
\end{align}
where $\theta_0$ and $\eta_0$ are free parameters. $\mathscr{T}$ is the time-period of the pump and $\tau$ is the instantaneous time. This protocol defines a closed loop in the $(\theta_2,\eta_2)$ plane, shown in \Fig{fig:Fig4}(a), which encloses the gap-closing point at $\theta_1=\theta_2$ and $\eta_1=\eta_2=0$. We compute the instantaneous quasi energy spectrum $\epsilon(\tau)$ using the formalism developed in Appendix~\ref{subsec:Appen_S_eigen_value_open} for an open network. The spectrum is shown in \Fig{fig:Fig4}(b), where the color indicates the spatial position of the phase eigenstates. Throughout the driving cycle, the bulk spectrum remains gapped. In contrast, edge states appear near $\epsilon=0$ and $\epsilon=\pi$ and exhibit spectral flow across the gaps. No edge-localized states are found between the bulk bands near $\epsilon=\pm \pi/2$.

We next quantify the charge pumped through the network. Within the driving protocol of \cref{eqn:driving_protocol}, we compute the instantaneous two-terminal scattering matrix of the network (see Appendix~\ref{sec:Appen_Rediffer_prod}):
\begin{align}
\mathcal S(\tau) =
\begin{pmatrix}
\mathcal R(\tau) & \mathcal T'(\tau) \\
\mathcal T(\tau) & \mathcal R'(\tau)
\end{pmatrix}.
\end{align}
During the driving cycle, the system remains gapped. As a result, the transmission amplitudes vanish, $\mathcal T(\tau) = \mathcal T'(\tau) = 0$, and the reflection amplitudes are unimodular, $\mathcal R(\tau) = e^{i\phi_{\mathcal R}(\tau)}$ due to unitarity of $\mathcal S$. The charge pumped during one driving cycle is given by the winding of the reflection phase $\phi_{\mathcal R}(\tau)$ \cite{Brouwer_1998}:
\begin{align}
Q = \frac{1}{2\pi}\bigl[\phi_{\mathcal R}(\mathscr{T}) - \phi_{\mathcal R}(0)\bigr].
\label{eq:qdef}
\end{align}
\cref{fig:Fig4}(c) shows the evolution of $\phi_{\mathcal R}(\tau)$ during the drive period. For $\epsilon$ in the bulk-gap close to $\epsilon=0$, the phase $\phi_{\mathcal R}(\tau)$ winds once, indicating that one unit of charge is pumped per cycle $(Q=1)$. This coincides with the spectral flow of the edge states observed in \cref{fig:Fig4}(b). In contrast, for $\epsilon$ in the bulk-gap close to $\epsilon=\pi/2$, where no spectral flow occurs, $\phi_{\mathcal R}$ does not undergo winding, and no charge is pumped into the network $(Q=0)$.

 Having discussed Thouless charge pumping in the network model, we further characterize its topological nature by computing the $q-\tau$ Chern number \cite{Thouless_pumping_83}, which provides an alternative interpretation of quantized charge pumping. Given the adiabatic protocol for the scattering network, we obtain the Chern number $C=1$, confirming that one unit of charge is pumped per cycle. This result is fully consistent with the pumping picture discussed above. Details of the formulation and numerical evaluation are presented in \cref{app:QTCHERN}.

We now study how a random dynamical phase \(\epsilon = k_{\rm F} L\) affects the pumped charge. Physically, this randomness represents fluctuations in the distances \(L\) between successive scatterers. We model this disorder by choosing \(\epsilon\) independently on each link from a uniform distribution, \(\epsilon \in [-W, W]\), where \(W\) sets the disorder strength. For each disorder realization, we compute the pumped charge (see \cref{eq:qdef,sec:Appen_Rediffer_prod}) and then average over realizations to obtain the disorder-averaged pumped charge \(\overline{Q}\). The dependence of \(\overline{Q}\) on \(W\) is shown in \cref{fig:Fig4}(d) for representative parameters. We find that \(\overline{Q}\) remains quantized at \(\overline{Q}=1\) up to a critical disorder strength \(W \approx \Delta_{\epsilon}\), where \(2\Delta_{\epsilon}\) is the phase gap around \(\epsilon=0\). Beyond this scale, the quantization breaks down. This shows that the quantized pumped charge is robust against dynamical phase disorder and is protected by the phase gap. Crucially, perfect periodicity of the scatterers is not required: quantization survives phase mixing as long as the disorder strength remains smaller than the phase-band gap.

\section{Lattice Realization}
\label{sec:latticereal}

\subsection{Model}

\begin{figure}
    \centering
    \includegraphics[width=1\linewidth]{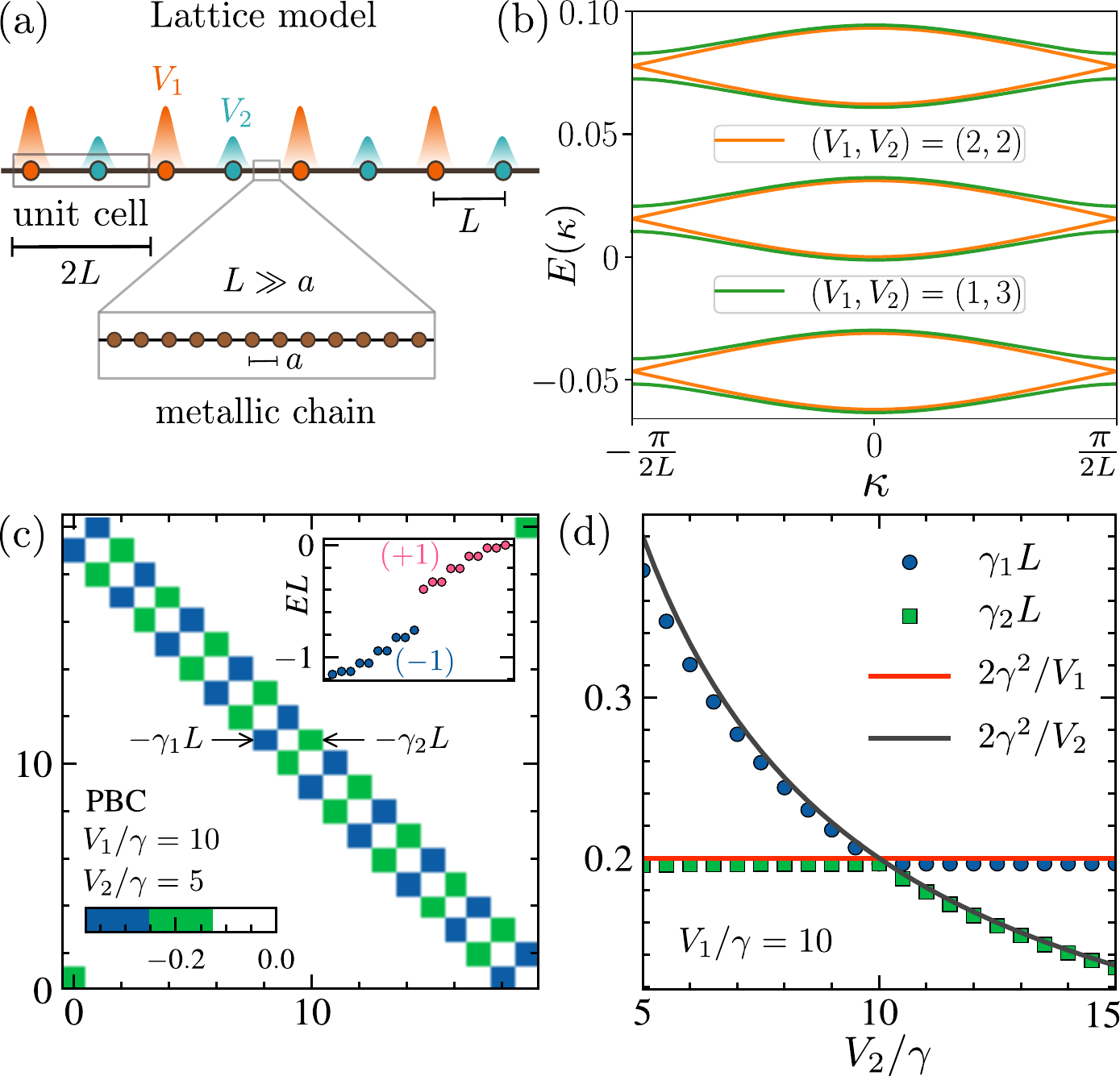} \caption{
    \textbf{Microscopic lattice model and its Wannierization.} 
(a) Tight-binding metallic chain with alternating on-site potentials \(V_1\) and \(V_2\), separated by distance $L$. 
(b) Bloch bands of the lattice model for \(L=100\) near zero energy. (c) Wannier Hamiltonian matrix (up to a chemical potential) for a pair of bands near zero energy, shown in the inset. 
(d) Values of $\gamma_1 L$ and $\gamma_2 L$ extracted from the Wannier Hamiltonian as a function of $V_2/\gamma$ with $V_1/\gamma$ held fixed. Analytical results for $\gamma_{1,2}$ are plotted for comparison. We use $\mathcal L = 1000$ and $L = 50$.
      }
\label{fig:Fig5}
\end{figure}

Having introduced the general properties of the SSH network model, we now present the microscopic lattice model that realizes such a network. We model the metallic wire of length $\mathcal{L}$ with a TB Hamiltonian for spinless electrons hopping between the nearest-neighbor sites as
\beq
H_{\text{metal}} = -\gamma \sum_{i=1}^{\mathcal{L}} c^\dagger_{i}c_{i+1} + \rm{h.c.} 
\label{eq:TB_metal}
\eeq
where $c_i^\dagger$ and $c_i$ are the fermion creation and annihilation operators at site $i$, respectively. We set the hopping amplitude $\gamma=1$ and the lattice spacing $a=1$. We further introduce a periodic array of impurities, modeled as alternating onsite potentials $V_1$ and $V_2$ at lattice sites separated by a distance $L \gg a $, forming a periodic chain of $\mathcal{N}$ superlattice unit cells, each of length $2L$ (see \Fig{fig:Fig5}(a)). The corresponding Hamiltonian is given by
\beq
H_{\text{defects}} = \sum_{j=1}^{\mathcal{N}} \Big( V_{1}c^\dagger_{(2j-1)L} c_{(2j-1)L} + V_{2}c^\dagger_{2jL} c_{2jL} \Big)
\label{eq:TB_on_site}
\eeq
The complete Hamiltonian is given by 
\beq
H= H_{\rm{metal}} + H_{\rm{defects}}
\label{eq:TBHam}
\eeq

For the translationally invariant superlattice, we apply Bloch's theorem and obtain the energy dispersion $E(\kappa)$ as a function of Bloch-momentum $\kappa$. Given that each unit cell has $\sim 2L$ sites, the system has a large number of bands within a reduced Brillouin zone. A numerical plot of mini-bands close to zero energy for the respective parameters $(V_1, V_2)$ is shown in \Fig{fig:Fig5}(b). Similarly to the phase bands of the network model, we find that when $V_1 \neq V_2$, the energy bands remain gapped. In contrast, for $V_1=V_2$ the gap closings occur at $2\kappa L= \pm \pi$. 

\subsection{Wannierization}

To obtain an effective tight-binding model for a pair of mini-bands, we \emph{Wannierize} the two bands closest to \(E=0\) and label them by indices \(\pm 1\) (see the inset of \cref{fig:Fig5}(c)). For simplicity, we impose PBC to eliminate boundary-related effects in the effective tight-binding model. Given that any neighboring pair of mini-bands produces approximately the same Wannierized Hamiltonian, the following discussion therefore applies to a generic pair of bands near zero energy. We first diagonalize the full Hamiltonian as \(H = U D U^\dagger\), where \(U_{in} = \langle i | \psi_n \rangle\). Here \(|i\rangle\) denotes the orbital basis, \(|\psi_n\rangle\) are the eigenstates of \(H\), and \(D\) is a diagonal matrix with elements \(D_{nn} = \varepsilon_n\).

To construct Wannier functions, we introduce a complex representation of the orbital positions \((x_j)\), \(z_j =  \exp\!\left( 2\pi i x_j / \mathcal L \right)\). The index \(j\) runs over all orbitals in the same order as used to construct the total Hamiltonian in ~\cref{eq:TB_metal}. The coordinates satisfy \(x_j \in [0,\mathcal L-1]\). This representation respects PBC along the \(x\)-direction. We then project the position operator onto the target subspace of eigenstates within the mini-band energy window, \(X_{p q} = \sum_j z_j \braket{\psi_p}{j} \braket{j}{\psi_q}\), where \(p,q\) label eigenstates with energy lying inside the pair of mini-bands $\pm 1$. Diagonalizing the projected position operator, \(X = V \Lambda V^\dagger\), yields the maximally localized Wannier orbitals (MLWOs) \(\ket{w_i} = \sum_n V_{n i} \ket{\psi_n}\). Here \(V_{n i} = \braket{\psi_n}{w_i}\), and \(\Lambda\) is a diagonal matrix with entries \(\Lambda_{n n} = \lambda_n\). The complex numbers \(\lambda_n\) encode the Wannier-center positions, given by \(x_n = \mathcal L \arg(\lambda_n) / 2 \pi\). Finally, the Wannierized tight-binding Hamiltonian is \([\mathcal H_{\mathrm{w}}]_{m n} = \matrixel{w_m}{H}{w_n}\).

The effective Hamiltonian \(\mathcal H_{\mathrm w}\) is shown in \cref{fig:Fig5}(c). For clarity, we subtract a constant ``chemical potential'' from \(\mathcal H_{\mathrm w}\). The effective tight-binding model features two alternating hopping amplitudes \(\gamma_1\) and \(\gamma_2\), giving it the same dimerized structure as the SSH model. Further, we examine how \(\gamma_1\) and \(\gamma_2\) depend on \(V_1\) and \(V_2\). In the strong-impurity limit \(V_{1,2}/\gamma \gg 1\), perturbation theory yields \(\gamma_1 \approx 2\gamma^2/(V_1 L)\) and \(\gamma_2 \approx 2\gamma^2/(V_2 L)\) (see Appendix~\ref{sec:pertth}). These analytical expressions agree very well with the numerically extracted hopping amplitudes, as shown in \cref{fig:Fig5}(d). Thus, the microscopic lattice model shows the emergence of SSH-like topological bands due to the effective impurity lattice.  
\subsection{Microscopic-to-network mapping}
\label{sec:mapping}

\begin{figure}
    \centering
    \includegraphics[width=1\linewidth]{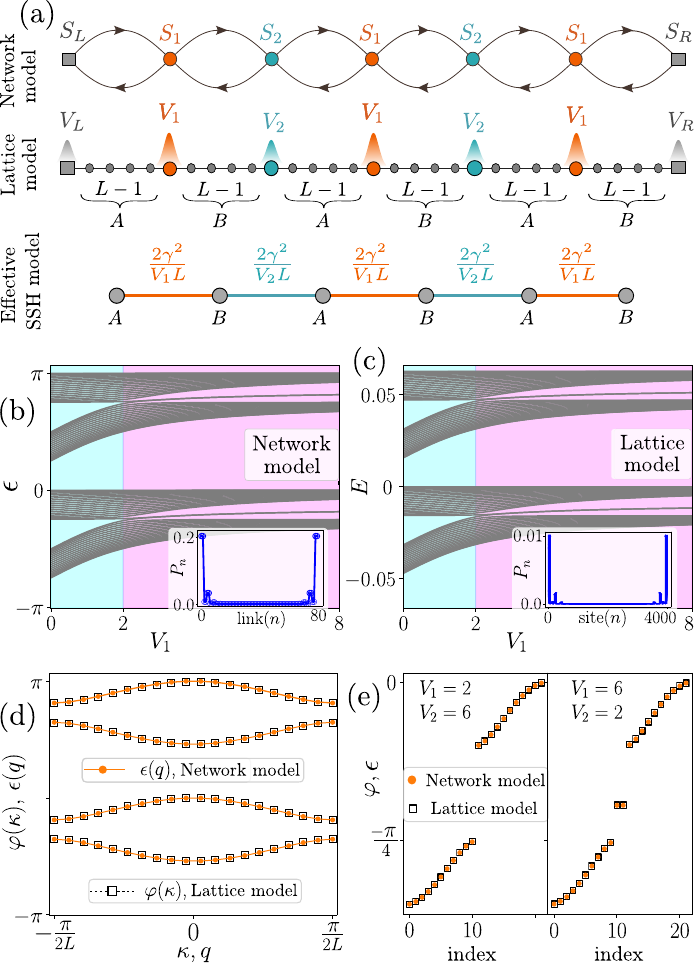} \caption{\textbf{Microscopic-to-network mapping.}  (a) Network model representation (top) of the microscopic lattice model (middle). The bulk impurities $V_1, V_2$ are described by scattering matrices $S_1, S_2$, while the boundary potentials $V_L$ and $V_R$ map onto terminal matrices $S_L$ and $S_R$, respectively. In the strong-impurity limit, the lattice model maps onto an effective SSH model (bottom). (b)-(c) Quasienergy spectrum  $\epsilon$ of the network model and energy spectrum $E$ of the lattice model close to zero energy for identical parameters $V_L=V_R=V_2=2$ with $20$ unit cells and $L=100$. Subgap edge modes appear for $V_1>V_2$ and are absent otherwise. Insets show the edge localization of the subgap modes. (d)-(e) Direct mapping between dynamical phase $\varphi$ and quasienergy $\epsilon$. (d) A plot of $\varphi(\kappa)$ and $\epsilon(q)$ as function of momentum $\kappa$ and $q$ respectively with periodic boundary conditions for $V_1=1$, $V_2=3$. (e) The same comparison with an open boundary for two different values of $(V_1, V_2)$. For clarity, the tight-binding indices are shifted so that the state near $E=0$ is labeled as index $0$. In both the periodic and open systems, the two spectra coincide, demonstrating a direct mapping between the two descriptions.}
    \label{fig:Fig6}
\end{figure}

In this section, we establish a direct correspondence between the network model and the microscopic lattice model description by constructing the network scattering matrix from the TB Hamiltonian. We use the scattering matrix derived for a TB chain with hopping $\gamma=1$ and on-site potential $V_{\alpha}$, for which the reflection $r_{\alpha}=r'_{\alpha} $ and transmission amplitudes $t_{\alpha}= t'_{\alpha}$ are given by \cite{datta1997electronic}
\beq
r_{\alpha}= \dfrac{V_{\alpha}}{2i\sin k_F- V_{\alpha}}, 
\quad
t_{\alpha}= \dfrac{2i\sin k_F }{2i\sin k_F- V_{\alpha}} 
\label{eq:rtfortb}
\eeq

The Fermi momentum $k_F$ is determined by the dispersion of the clean chain, $E(k)=-2\cos k$. These amplitudes establish a one-to-one correspondence between the microscopic impurity strength $V_{\alpha}$ and the parameters of the scattering matrix $S_{\alpha}$ of the network. Hence, the network description is fully determined by the lattice Hamiltonian. This mapping is illustrated schematically in \cref{fig:Fig6}(a). The top panel presents the network model with alternating scattering matrices $S_1$ and $S_2$, obtained from \cref{eq:rtfortb} for impurities $V_1$ and $V_2$. The corresponding lattice model is shown in the middle panel. In the strong-impurity limit, the microscopic lattice model maps onto an effective SSH model with alternating couplings $\gamma_1=2\gamma^2/(V_1 L)$ and $\gamma_2=2\gamma^2/(V_2 L)$ as illustrated in the bottom panel of \cref{fig:Fig6}(a).
A crucial step in mapping the lattice model onto the network model is implementing equivalent boundary conditions in both descriptions. Alternatively, the same boundary physics can be examined at an interface between a trivial and a topological network, where the boundary arises naturally rather than being imposed by hand. In what follows, we adopt the former approach, explicitly fixing boundary conditions, and refer the reader to \cref{app:interface} for a discussion of the interface formulation. 

In the lattice model, we consider general on-site potentials $V_L$ and $V_R$ at the left and right ends of the chain. Since the termination of the lattice enforces no propagating states outside the system, under the network mapping, this translates into modified terminal scattering matrices $S_L$ and $S_R$ with reflection amplitudes $r_{L,R}$ given by
\beq
r_{L,R}= -\dfrac{V_{L,R} + e^{ik_F}}{V_{L,R} + e^{-ik_F}}, 
\label{eq:rtforedge}
\eeq
with perfect reflection $|r_{L,R}|=1$ and vanishing transmission $|t_{L,R}|=0$ (see \cref{app:rtends} for details). Having fixed the mapping, we now compare the resulting spectrum in the two descriptions. For concreteness, we set $V_L=V_R=V_2$. The \cref{fig:Fig6}(b) presents the quasienergy spectrum $\epsilon$ of the network model as a function of $V_1$ and a fixed value of $V_2=2$. For the same parameters, \cref{fig:Fig6}(c) shows the minibands energy spectrum $E$ obtained from the microscopic lattice model near zero energy. In both descriptions, when $V_1> V_2$, the spectrum supports two degenerate subgap modes separated from the bulk spectrum, while for $V_1<V_2$, no subgap modes are present. Furthermore, the localization probability density $P_n$ of these modes is strongly localized at the ends of the chain, as illustrated in the insets. Interestingly, the two spectra exhibit identical dispersions, and up to a scale factor, the results are identical. This close similarity indicates an underlying relation between the lattice energy $E$ and the network quasienergy $\epsilon$, determined by the dynamical phase accumulated during ballistic propagation between successive scatterers.

To make this relation explicit, we evaluate the group velocity $v_g=\partial_k E(k)=2\sin k$. We fix the Fermi energy at $E(k_F)=0$, which corresponds to $k_F=\pi/2$ and $v_g=2$. An electron with energy $E$ propagates ballistically along the metallic chain between successive scatterers over a time $L/v_g$. During this propagation, it acquires a dynamical phase $\varphi = E L/(\hbar v_g)$, which coincides with the network quasienergy $\epsilon$ up to an overall constant phase shift.

For energies close to $E=0$, we compare the dynamical phase $\varphi$ of the microscopic lattice model with the quasienergy $\epsilon$ of the network model, as shown in \cref{fig:Fig6}(d)-(e) for $L \gg a$. With periodic boundary conditions, the bulk bands of $\varphi$ and $\epsilon$ overlap up to an overall phase shift [see \cref{fig:Fig6}(d)]. The same agreement holds for open boundary conditions [see \cref{fig:Fig6}(e)], where both bulk bands and edge modes coincide. This shows that the network model faithfully reproduces both the spectral and topological properties of the underlying lattice model in the low-energy regime and remains fully consistent with the effective SSH model description in the strong-impurity limit. It is important to note that the microscopic lattice model [see \cref{eq:TBHam}] belongs to class AI of the ten-fold classification \cite{Kitaev_AIP_2009, Agarwala_PRL_2017}, which does not host a topological phase in one dimension. However, impurity physics induces an effective sub-lattice symmetry in the minibands, leading to an emergent BDI symmetry which results in protected SSH physics in one dimension (see Appendix~\ref{sec:pertth}). 

Achieving exact quantitative agreement requires an even larger separation between scatterers, which makes lattice-model simulations computationally expensive. For open boundary conditions, the tight-binding approach requires diagonalizing an $(2 \mathcal N L)\times(2 \mathcal N L)$ scattering matrix, whereas the network model only involves an $2\mathcal N\times 2\mathcal N$ unitary matrix to obtain the phase bands, with $\mathcal N$ the number of unit cells. Similarly, for periodic boundary conditions, the lattice model requires diagonalizing a $(2L)\times(2L)$ Hamiltonian, while the network model reduces the problem to a $4\times4$ unitary matrix.
Hence, the network model offers a clear and substantial numerical advantage.

\section{Potential realizations of SSH Network: Array of Quantum Point Contacts}
\label{sec:realisation}

\begin{figure}[ht]
    \centering
    \includegraphics[width=0.9\linewidth]{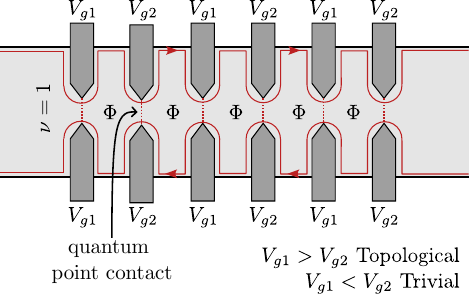}
    \caption{\textbf{Quantum point contact (QPC) based realization of the network model}. 
   Arrays of QPCs with dimerized gate potential on a Hall bar geometry in the Lowest Landau level.
    }
    \label{fig:Fig_4}
\end{figure}

Having discussed a microscopic lattice model and its direct mapping to the SSH network, we now discuss another physical avenue for realizing such a network using a quantum point contact (QPC). QPC in the Hall bars have been of immense interest for studying scattering between various integer and fractional quantum Hall edge states and statistics of their quasiparticles in two-dimensional electron gases and graphene systems \cite{Carrega_2021}. We consider a quantum Hall bar in the integer quantum Hall regime under a uniform magnetic field \(B\), with cyclotron frequency \(\omega_c = eB/m\), where \(e\) and \(m\) are the electron charge and mass. Quantum point contacts (QPCs) with alternating gate voltages \(V_{g1}\) and \(V_{g2}\) are patterned along the Hall bar, as shown in \cref{fig:Fig_4}(b).

To model this system, we describe each QPC by its scattering matrix and map the array of QPCs onto a network model. Each QPC is modeled as a saddle-point potential \cite{Fertig_1987},
\begin{align}
V_{\mathrm{SP},\alpha} = -U_x x^2 + U_y y^2 + V_{g\alpha} ,
\end{align}
where \(V_{g\alpha}\) denotes the gate voltage applied to the QPC, $\alpha =1,2$ labels the two alternating gate voltages.

The transmission probability at the Fermi energy \(E_F\) is
\begin{align}
T_{\alpha}(E_F) &= \frac{1}{1 + \exp(-\pi \eta_{\alpha}(E_F))}, \\
\eta_{\alpha}(E_F) &= \frac{E_F - (n + 1/2)E_2 - V_{g\alpha}}{E_1}.
\end{align}
For a weak saddle-point potential, \(U_x,U_y \ll m\omega_c^2\), we have
\(E_2 \approx \hbar\omega_c\) and
\(E_1 \approx \sqrt{U_x U_y}/(m\omega_c)\).
Restricting to the lowest Landau level (\(n=0\)), transmission is controlled by gate voltage \(V_{g\alpha}\):
\(T \approx 1\) for \(V_{g\alpha} \ll E_F - \hbar\omega_c/2\), and
\(T \approx 0\) for \(V_{g\alpha} \gg E_F - \hbar\omega_c/2\). We can then model the corresponding scattering matrix of the QPC following \cref{General_s_mat}
\begin{align}
S_{\alpha} =
\begin{pmatrix}
\sqrt{1-T_{\alpha}} & i\sqrt{T_{\alpha}} \\
i\sqrt{T_{\alpha}} & \sqrt{1-T_{\alpha}}
\end{pmatrix}.
\label{eq:qpcS}
\end{align}

The phase accumulated along each link includes the Aharonov--Bohm contribution,
\(\epsilon \to \epsilon + \pi \Phi/\Phi_0\),
where \(\Phi_0 = h/e\) is the flux quantum and \(\Phi/\Phi_0\) is the number of flux quanta enclosed between neighboring QPCs.
The link phase can be tuned by an external magnetic field, without changing the distance between the QPC gates, which is experimentally convenient.

Comparing this scattering matrix \cref{eq:qpcS} with the general parametrization in Eq.~\eqref{General_s_mat}, we identify $\phi_{\alpha}=0$ and $\theta_{\alpha}=\sin^{-1}(\sqrt{T_{\alpha}})$. For this system, $V_{g1} > V_{g2}$ leads to a topological phase, whereas $V_{g1} < V_{g2}$ renders the network of QPC array in the trivial limit. Hence, with such QPC array patterning, one can realize an effective SSH network along the edge.

Further, the SSH-network can also be realized on the edge of quantum spin Hall insulator such as seen in HgTe/CdTe \cite{Konig_Science_2007a}, InAs/GaSb \cite{knez2011evidence} quantum well structures, or two dimensional systems such as monolayer WSe$_2$ \cite{Chen_2018} and WTe$_2$ \cite{Wu_2018}, by decorating the edge with alternating magnetic impurities which causes scattering between the spin-flavoured edge modes. Then, tuning the magnetic impurity strengths allows one to go from a trivial to a topological phase.

\section{outlook}
\label{sec:outlook}

Advances in engineering of defects and dopants in mesoscopic systems \cite{Fuechsle2012NatNano, Gorman2019NatNano} have been of immense interest given the ability to control systems at the atomic scale. The realization of topological phases across various platforms has been a recurrent theme in modern condensed matter physics, partly driven by interest in topological quantum computing. In this work, we have shown that designing impurity lattices can induce such topological physics, leading to the realization of the SSH network. The latter network can be completely characterized within a scattering matrix formalism, which can capture a host of different systems where scattering modes scatter coherently from impurities. 

We show that such scattering networks realize trivial and topological phases where the Fermi energy and inter-impurity distance can lead to insulating and conducting transport, as tunability among the quasienergy bands is engineered by impurity strengths. In particular, we show that the network has two `atomic' limits which are trivial and topological in character depending on the existence of edge modes. Moreover, the system realizes the tell-tale signatures of the topological phases, such as Thouless-charge pump and stability to disorder. We further show that in a microscopic lattice model, the scattering network can be exactly realized, where the energy band structure description matches the description from the scattering network. We further motivated that such physics is applicable to other material platforms, such as an array of QPCs. The same physics may be relevant in the recently discovered one-dimensional Moire systems where carbon-nanotubes see a slow-wavelength periodic potential \cite{ArroyoGascon2020NanoLett, Bi2025arXiv,Zhao2020PRL}. 

In our study, we have focused exclusively on the ballistic regime, where no incoherent scattering processes are included. It will be interesting to systematically include such processes to note if any temperature-dominated diffusive transport exists in such a system. Additionally, we have focused on non-interacting systems; in the presence of electron-electron interactions, each of the metallic regions is expected to convert into Luttinger liquids. It will be interesting to note how the topological and trivial phases of the scattering network vary due to the finite Luttinger parameter \cite{chen2025tunable, Wang2022Nature,HuLian2024PRB}.

Another interesting future direction will be to explore this physics in higher dimensions. Exact microscopic mappings between lattice models and effective Chalker-Coddington networks \cite{chalker1988percolation, ChalkerPRL2020Floquet,Chalker1996networkmodels} will be interesting to explore and to show if higher-order phases can be realized in them. Such exact correspondences may be of immediate interest both from theoretical directions and from networks of edge states realized in twisted systems \cite{Pasek2014Networkmodels,  Kundu_PRB_2024, Bilayergraphenenetwork2021}. The role of interactions, or fractional quantum Hall edges, in equilibration and transport for both electrical and heat transport will be of immense interest \cite{Dolev2008Nature, Banerjee2018Nature}.

\section{Acknowledgments}

We acknowledge fruitful discussions with Sumathi Rao and Diptarka Das. AA acknowledges funding from projects IITK/PHY/2022010 and IITK/PHY/2022011.  RK acknowledges support from the FARE program at IIT Kanpur. RK acknowledges support from the FARE program at IIT Kanpur and the NPDF scheme of ANRF Grant No. PDF/2025/002176. RS acknowledges funding from the IIT Kanpur Institute Fellowship.

\appendix 

\section{Symmetries of the scattering matrix}
\label{scatsymm}
\subsection{Time reversal symmetry}
 In the case of spinless fermions, the time reversal operation is just conjugation: \( \mathcal T = K\), (\(K\) is conjugation). The time reversal operation also interchanges the incoming and outgoing modes. The action of the time-reversal operation on incoming ($\bm a$) and outgoing modes ($\bm b$) is defined as 
\begin{align}
	\label{eq:app:trs1}
	&\mathcal T \bm a = \bm b^*,\quad
	\mathcal T \bm b  = \bm a^* 
\end{align}
Action of $\mathcal T$ on the scattering matrix $S$ is
\begin{align}
	\label{eq:app:sminus}
	\mathcal T S \mathcal T^{-1} = S^*
\end{align}

The scattering matrix \( S \) relates incoming and outgoing modes
\begin{align}
	\label{eq:app:scat1}
	\bm b = S \bm a
\end{align}

By acting time reversal operation on Eq.~\eqref{eq:app:scat1} we obtain
\begin{align}
	\label{eq:app:scat2}
	&\mathcal T \bm b = \mathcal T S \mathcal T^{-1} \mathcal T
	\bm a \\
	\Rightarrow &\bm a^* = S^* \bm b^* \\
	\label{eq:app:scat3}
	\Rightarrow &\bm b = S^{T} \bm a 
\end{align}
To arrive at \cref{eq:app:scat3} from \cref{eq:app:scat2} we use
\cref{eq:app:sminus} and \cref{eq:app:trs1}. Now, if the scattering matrix is
invariant under time reversal, then \cref{eq:app:scat1} and
\cref{eq:app:scat3} together imply that \( S = S^{T} \), i.e., for time reverse symmetry obeying scattering process, the scattering matrix has to be symmetric.
\subsection{Inversion symmetry}
We derive the constraints imposed by inversion symmetry on the scattering matrix of a single-mode wire.
For a single channel, the incoming and outgoing amplitudes,
\(\bm a = (a_1,a_2)^T\) and \(\bm b = (b_1,b_2)^T\), are related by
\begin{align}
    \bm b = S \bm a , \qquad
    S =
    \begin{pmatrix}
        r & t' \\
        t & r'
    \end{pmatrix},
\end{align}
where \(r\) (\(r'\)) denotes the reflection amplitude for incidence from the left (right), and
\(t\) (\(t'\)) denotes the corresponding transmission amplitude.

Inversion symmetry exchanges left- and right-moving modes,
\begin{align}
    a_1 \leftrightarrow a_2, \qquad
    b_1 \leftrightarrow b_2 ,
\end{align}
which we represent by the inversion operator \(\mathcal I\),
\begin{align}
    \mathcal I \bm a = \sigma_x \bm a, \qquad
    \mathcal I \bm b = \sigma_x \bm b,
\end{align}
with \(\sigma_x\) the Pauli matrix.

Invariance of the scatterer under inversion implies
\begin{align}
    S = \mathcal I S \mathcal I^{-1}.
\end{align}
Applying inversion to the scattering relation \(\bm b = S \bm a\) gives
\begin{align}
    \sigma_x \bm b = S \sigma_x \bm a .
\end{align}
Using \(\bm b = S \bm a\), this leads to the constraint
\begin{align}
    S = \sigma_x S \sigma_x .
\end{align}
Evaluating this condition yields
\begin{align}
    r = r', \qquad t = t'.
\end{align}
Thus, inversion symmetry enforces identical reflection and transmission amplitudes for left and right incidence, and the scattering matrix reduces to
\begin{align}
    S =
    \begin{pmatrix}
        r & t \\
        t & r
    \end{pmatrix}.
\end{align}

\section{Transmission for translationally symmetric network}
\label{transmission}

Here we specify the parameters of the local scattering matrices $S_{\alpha}$, by modeling each impurity of strength $V_{\alpha}$ as a one-dimensional 
Dirac-delta potential scatterer \cite{markos2008wave}. The associated scattering matrix is given by
\begin{equation}
S_{\alpha}=
\begin{pmatrix}
r_{\alpha} & t'_{\alpha} \\
t_{\alpha} & r'_{\alpha}
\label{eqn:Appen_s_mat}
\end{pmatrix},
\end{equation}
with
\begin{equation}
r_{\alpha}=r'_{\alpha}=\frac{k_{\alpha}}{ik-k_{\alpha}}, \qquad
t_{\alpha}=t'_{\alpha}=\frac{ik_{\alpha}}{ik-k_{\alpha}}.
\end{equation}
where $k$ is the incident wave vector and the parameter $k_{\al}$ characterizes the strength of the scatterer with $k_{\al}= m V_\alpha/{\hbar^2}$. In all numerical calculations, we set $m=1$ and $\hbar=1$. 

We use the transfer-matrix formalism used to compute the transmission \(T\) through a scattering network \cite{markos2008wave}. For a scatterer having index \(\alpha\), the transfer matrix \(M_\alpha\) relates the wave-function amplitudes on the left side, \((a_\alpha, b_\alpha)\), to those on the right side, \((\tilde b_\alpha, \tilde a_\alpha)\),
\begin{equation}
\begin{pmatrix}
\tilde{b}_{\alpha}\\
\tilde{a}_{\alpha}
\end{pmatrix}
=
M_{\alpha}
\begin{pmatrix}
a_{\alpha}\\
b_{\alpha}
\end{pmatrix}.
\end{equation}
The transfer matrix can be written in terms of the scattering matrix elements \(S_\alpha\) as
\begin{equation}
M_{\alpha}
=\frac{1}{t'_{\alpha}}
\begin{pmatrix}
t_{\alpha}t'_{\alpha}-r_{\alpha}r'_{\alpha} & r'_{\alpha}\\
-r_{\alpha} & 1
\end{pmatrix}.
\end{equation}

Between two successive scatterers \(\alpha\) and \(\beta\), the electron propagates freely and accumulates a dynamical phase \(\epsilon\). This process is described by a ``propagation matrix" \(P_{\alpha,\beta}\), which relates the outgoing amplitudes from scatterer \(\alpha\) to the incoming amplitudes at scatterer \(\beta\),
\begin{equation}
\begin{pmatrix}
a_{\beta}\\
b_{\beta}
\end{pmatrix}
=
P_{\alpha,\beta}
\begin{pmatrix}
\tilde{b}_{\alpha}\\
\tilde{a}_{\alpha}
\end{pmatrix},
\end{equation}
with
\begin{equation}
P_{\alpha,\beta}
=
\begin{pmatrix}
e^{i\epsilon} & 0\\
0 & e^{-i\epsilon}
\end{pmatrix}.
\end{equation}

The transfer matrix for a single unit cell of the network is obtained by combining scattering and propagation processes,
\begin{equation}
M_{\rm cell}=P_{2,3}M_2P_{1,2}M_1.
\end{equation}

For a network consisting of \(N\) identical unit cells, the total transfer matrix is given by the ordered product
$M_{\rm open}=M_{\rm cell}^N$.
Direct evaluation of this product can lead to numerical instabilities for large \(N\). For a translationally invariant system, the transmission can instead be computed analytically using Chebyshev’s identity,
\begin{equation}
T=\frac{1}{1+|M_{\rm cell}|^2\,\dfrac{\sin^2(N q L)}{\sin^2(qL)}}.
\end{equation}
Here \(qL\) is determined from the eigenvalues of \(M_{\rm cell}\), which are \(e^{\pm i q L}\). In transmission band gaps, \(q\) becomes complex, leading to an exponential suppression of transmission. Given any value of $V_1, V_2, k, \epsilon$, we calculated $T$, which was shown in \Fig{fig:Fig2} in the main text.

\section{Role of the phase \texorpdfstring{$\phi_{1,2}$}{} in the quasienergy spectrum}
\label{sec:finiteshift}

\begin{figure}
    \centering
    \includegraphics[width=1\linewidth]{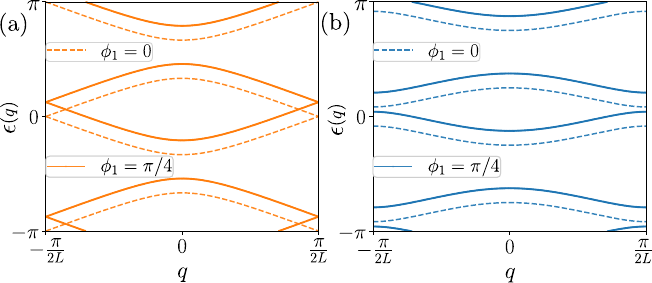}
    \caption{\textbf{Role of finite phases $\phi_{1,2}$}. Quasienergy spectrum for finite phase $\phi_{1}$ with $\phi_{2}=0$. (a) At the critical point $\theta_{1}=\theta_{2}= \pi/3$. (b) Away from criticality ($\theta_{1}= \pi/6,\theta_{2}=\pi/3$). Finite $\phi_{1}$ produces an overall shift of the quasienergy bands without modifying their dispersion.
    }
    \label{fig:finiteshift}
\end{figure}

In the main text (see \cref{subsec:Quasibands}), we set the phases $\phi_{1}=\phi_{2}=0$ without loss of generality, since they result in an overall phase shift of the spectrum. Here, we explicitly verify the statement considering the finite value of $\phi_{1}$ and set $\phi_{2}=0$; the same conclusions hold for any other choice of $\phi_{1,2}$. As shown in Fig.~\ref{fig:finiteshift}, a nonzero $\phi_{1}$ leads to a constant shift of the entire spectrum. In particular, the gap-closing condition at $\theta_{1}=\theta_{2}$ remains intact for arbitrary $\phi_{1,2} $, and no gap is opened at the critical point (panel~(a)). Away from criticality ($\theta_{1}\neq\theta_{2}$), finite $\phi_{1,2}$ again only shift the bands without modifying the existing gaps (panel~(b)).

\section{Quasienergy spectrum for an open network}
\label{subsec:Appen_S_eigen_value_open}

\begin{figure}
    \centering
    \includegraphics[width=0.9\linewidth]{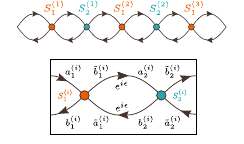}
    \caption{\textbf{Open network and scattering conventions}. Schematic of a finite open scattering network with alternately placed scatterers of type $S_{1}$, $S_{2}$, arranged with nodes labeled by $i$. The lower panel shows the incoming and outgoing conventions for a generic scatterer pair $S^{(i)}_{1}$ and $S^{(i)}_{2}$. 
    }
    \label{fig:Appen_open_net}
\end{figure}

We consider an open network composed of two types of scattering nodes, characterized by the scattering matrices $S_1$ and $S_2$ with $N_1$ nodes of type $S_1$ and $N_2$ nodes of type $S_2$, where $N_2= N_1-1$. We use the superscript $(i)$ to label the nodes as shown in \Fig{fig:Appen_open_net} where $S^{(i)}_{\alpha}$ represents the scattering matrix corresponding to the scatterer $\alpha$ having the index $(i)$ in the network. The corresponding incoming and outgoing amplitudes are 
$(a^{(i)}_{\alpha}, \tilde{a}^{(i)}_{\alpha})$ and $(b^{(i)}_{\alpha}, \tilde{b}^{(i)}_{\alpha})$, respectively. Using the scattering matrices of the individual nodes, the full scattering matrix of the network $S_{\rm{open}}$ is given by the ordered direct sum
\begin{align}
    S_{\rm{open}}= \left[ \bigoplus_{i=1}^{N_2} S^{(i)}_{1} \oplus S^{(i)}_{2} \right] \oplus S_1^{(N_1)}
    \label{S_open_tensor}
\end{align}
The dimension of the unitary matrix $S_{\rm{open}}$ is $N_d=2(N_1+N_2)$ where $N_1+N_2$ is the total number of nodes in the network. The $S_{\rm{open}}$ acts on all the incoming amplitudes $\bm{a}$ and relates to all the outgoing amplitudes $\bm{b}$ of the network via
\begin{align}
    \bm{b}= S_{\rm{open}} \bm{a},
    \label{eqn:S_open_relaion}
\end{align}
where 
\begin{align}
\bm{a}&= \bigoplus_{i=1}^{N_2} \bigg[\bigoplus_{\alpha=1,2}(a^{(i)}_{\alpha}, \tilde{a}^{(i)}_{\alpha})^{\rm{T}}\bigg] \oplus (a^{(N_1)}_{1},\tilde{a}^{(N_1)}_{1})^{\rm{T}} \nonumber \\
\bm{b}&= \bigoplus_{i=1}^{N_2} \bigg[\bigoplus_{\alpha=1,2}(b^{(i)}_{\alpha}, \tilde{b}^{(i)}_{\alpha})^{\rm{T}}\bigg] \oplus (b^{(N_1)}_{1},\tilde{b}^{(N_1)}_{1})^{\rm{T}}
\end{align}

In the bulk of the network, the outgoing modes and the incoming mode amplitudes are related by \cref{eq:links}. 
We model the edges so that modes are perfectly reflected from them and are related by the dynamical phase. This physically corresponds to an infinite potential at the edge.
\begin{align}
    a^{(1)}_{1}&= e^{i\epsilon} b^{(1)}_{1} \\
    \tilde{a}^{(N_1)}_{1}&= e^{i\epsilon} \tilde{b}^{(N_1)}_{1}
\end{align}
Using the dynamical phase relations, the incoming amplitudes $\bm{a}$ are expressed in terms of the outgoing amplitudes $\bm{b}$ as
\beq
\bm{a}= e^{i \epsilon} \mathscr{P}_{\rm{open}} \bm{b}
\eeq
where $\mathscr{P}_{\rm{open}}$ is a permutation matrix that reorders the outgoing amplitudes in the dynamical phase equations to ensure a consistent basis on both sides of the scattering equation (\cref{eqn:S_open_relaion}). Substituting it yields the eigenvalue equation
\beq
S_{\rm{open}} \mathscr{P}_{\rm{open}} \bm{b} = e^{-i \epsilon} \bm{b}. 
\label{eqn:open_s_mat_eign_val}
\eeq
By solving the eigen-problem in \cref{eqn:open_s_mat_eign_val}, we numerically obtain the quasi energy spectrum of the open network $\epsilon$ as a function of $\theta_1$ shown in the main text (\cref{fig:Fig2}(c)).

\section{Transmission for disordered network}
\label{sec:Appen_Rediffer_prod}

\begin{figure}
    \centering
    \includegraphics[width=0.9\linewidth]{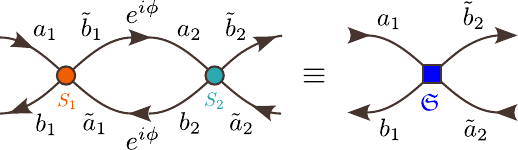}
    \caption{\textbf{Composing two scattering matrices}. The two scattering matrices $S_1$ and $S_2$, with associated incoming and outgoing amplitudes, are combined into an effective scattering matrix $\mathfrak{S}$, which relates the external amplitudes of the scatterers. }
    \label{fig:Compos_two_S_mat}
\end{figure}

Chebyshev identities provide an efficient way to compute transmission in translationally invariant systems. However, for disordered networks, one must evaluate the ordered product of transfer matrices for the entire system. Since transfer matrices are not unitary, this procedure is numerically unstable. Several methods have been developed to address this issue \cite{MacKinnon1983, Kramer1993}. Here, we adopt a scattering-matrix-based approach, which remains numerically stable and is therefore well suited for disordered systems.

We begin by deriving the composition rule for two scattering matrices \(S_1\) and \(S_2\) connected by a link, as shown in \Fig{fig:Compos_two_S_mat}. Each scatterer relates incoming and outgoing amplitudes via
\begin{align}
\begin{pmatrix}
b_{1}\\
\tilde{b}_{1}
\end{pmatrix}
&=
\begin{pmatrix}
r_1 & t_1'\\
t_1 & r_1'
\end{pmatrix}
\begin{pmatrix}
a_{1}\\
\tilde{a}_{1}
\end{pmatrix}, \\
\begin{pmatrix}
b_{2}\\
\tilde{b}_{2}
\end{pmatrix}
&=
\begin{pmatrix}
r_2 & t_2'\\
t_2 & r_2'
\end{pmatrix}
\begin{pmatrix}
a_{2}\\
\tilde{a}_{2}
\end{pmatrix}.
\end{align}

The link between \(S_1\) and \(S_2\) introduces a dynamical phase \(\epsilon\). Eliminating the internal amplitudes, the two scatterers can be combined into an effective scattering matrix \(\mathfrak{S}\), defined by
\begin{align}
\begin{pmatrix}
b_{1}\\
\tilde{b}_{2}
\end{pmatrix}
=
\mathfrak{S}
\begin{pmatrix}
a_{1}\\
\tilde{a}_{2}
\end{pmatrix},
\end{align}
with
\begin{align}
\mathfrak{S} =
\begin{pmatrix}
r_1 + \dfrac{r_2 t_1 t_1' e^{2 i \epsilon}}{1 - r_1' r_2 e^{2 i \epsilon}} &
\dfrac{t_1' t_2' e^{i \epsilon}}{1 - r_1' r_2 e^{2 i \epsilon}} \\[2.5ex]
\dfrac{t_1 t_2 e^{i \epsilon}}{1 - r_1' r_2 e^{2 i \epsilon}} &
r_2' + \dfrac{r_1' t_2 t_2' e^{2 i \epsilon}}{1 - r_1' r_2 e^{2 i \epsilon}}
\end{pmatrix}.
\label{eq:scombine}
\end{align}
We denote this composition by \(\mathfrak{S} = S_1 \star S_2\), where \(\star\) is a binary operation that combines two scattering matrices separated by a link.

Using this composition rule, the full scattering matrix of the open network in \Fig{fig:Appen_open_net} can be constructed by successively combining the scattering matrices of individual elements. The resulting effective scattering matrix \(\mathcal{S}\) relates the incoming amplitudes at the two open ends of the network, \((a^{(1)}_{1}, \tilde{a}^{(N_1)}_{1})^{\mathrm T}\), to the corresponding outgoing amplitudes, \((b^{(1)}_{1}, \tilde{b}^{(N_1)}_{1})^{\mathrm T}\),
\begin{equation}
\mathcal{S}
=
\left[
\mathop{\bigstar}_{i=1}^{N_2}
(S^{(i)}_{1} \star S^{(i)}_{2})
\right]
\star
S^{(N_1)}_{1}.
\label{eqn:Appn_Redh_prod_chain}
\end{equation}

The effective matrix \(\mathcal{S}\) is a \(2\times2\) unitary matrix,
\begin{align}
\mathcal{S} =
\begin{pmatrix}
\mathcal{R} & \mathcal{T}'\\
\mathcal{T} & \mathcal{R}'
\end{pmatrix},
\end{align}
where \(\mathcal{R}\) and \(\mathcal{R}'\) are the reflection amplitudes for waves incident from the left and right, respectively, and \(\mathcal{T}\) and \(\mathcal{T}'\) are the corresponding transmission amplitudes. This method is used in the main text to compute transport through disordered networks in \cref{sec:topological_network}, and to analyze charge pumping in \cref{sec:Thouless_pump}.

\section{\texorpdfstring{$q-\tau$}{} Chern number}
\label{app:QTCHERN}

In this appendix, we compute the $q-\tau$ Chern number for the network model. This invariant provides a direct measure of the quantized charge pumping. It is defined over the two-dimensional parameter space spanned by the momentum $q$ and the adiabatic parameter $\tau$. 
\begin{align}
C= \frac{-i}{2\pi } \int_{0} ^{T} d\tau  \int_{-\pi} ^{\pi} dq \Omega(q,\tau)
\end{align}
with 
\begin{align}
\Omega(q, \tau)= \partial_q  (\bm{b}^{\dagger}(q,\tau) \partial_{\tau}  \bm{b}(q,\tau) )- \partial_\tau (\bm{b}^{\dagger}(q,\tau) \partial_{q}  \bm{b}(q,\tau))
\end{align}
Instantaneous eigenvectors $\bm{b}(q,\tau)$ are obtained by diagonalizing the eigenvalue equation \cref{eq:eigen} defined in the main text, evaluated for the time-dependent scattering matrix $\tilde{S}_{\alpha}$ specified for the adiabatic protocol \cref{eqn:driving_protocol}.
Using the formulation developed by Fukui and Hatsugai \cite{FukuiHatsugaiSuzuki2005}, we numerically compute the Chern number $C=1$ for the present network model. This confirms that the charge pumped during one driving cycle is quantized.

\section{Effective Hamiltonian from the tight-binding model using perturbation theory}
\label{sec:pertth}

Consider the Hamiltonian [see \cref{eq:TBHam}] where under PBC, the total number of sites is ${\cal L} =2NL$. Defining a unit cell with $2L$ number of sites  and redefining,
\begin{align}
c^{\dagger}_{2nL+1}  \rightarrow  & d^{\dagger}_{nA} \qquad c^{\dagger}_{2nL+\ell+1}  \rightarrow   c^{\dagger}_{\ell,nA}  \\
c^{\dagger}_{(2n+1)L+1}  \rightarrow  & d^{\dagger}_{nB} \qquad c^{\dagger}_{(2n+1)L+\ell+1}  \rightarrow   c^{\dagger}_{\ell,nB}
\end{align}
where $n\in \{0, \ldots, N-1\}$ and 
$\ell\in \{1, \ldots, L-1\}$. The Hamiltonian 
can be written as 
\begin{align}    
H_{\text{metal}} &= \sum_{\alpha=A,B} \sum_{n=0}^{N-1}\sum_{i=1}^{L-1} -\gamma \Big( c^\dagger_{i,n\alpha} c_{i+1,n\alpha} + \text{h.c.} \Big)\\
H_{\text{imp}} &= \sum_{n=0}^{N-1} \Big( V_1 d^\dagger_{nA} d_{nA}  + V_2 d^\dagger_{nB} d_{nB} \Big)
\end{align}
\begin{dmath}
H_{\text{hyb}} =  \sum_{n} -\gamma \Big((d^\dagger_{nA}c_{1,nA} + \text{h.c.}) + (d^\dagger_{nB}c_{1,nB} + \text{h.c.}) \Big)
+ \sum_{n} -\gamma \Big((d^\dagger_{nA}c_{L-1,n-1 B} + \text{h.c.})+ (d^\dagger_{nB}c_{L-1,nA} + \text{h.c.}) \Big)
\end{dmath}

Diagonalizing $H_{\text{metal}}$ separately, 
\begin{align}
H_{\text{metal}} = \sum_{\alpha=A,B}
\sum_{n=0}^{N-1}\sum_{k}   E(k) c^\dagger_{k,n\alpha} c_{k,n\alpha} 
\end{align}
where $E(k) = - 2\gamma \cos(k)$ where $k = \frac{2\pi m}{L}$ and $m=\{0,\ldots, L-1\}$. Given a value of $L$, the gaps are of the order $\sim 2\gamma /L$, which are degenerate for each value of $n$. 

We are interested in projecting the perturbation $V = H_{\text{hyb}}$ into the low-energy subspace of $H_o = H_{\text{metal}} + H_{\text{imp}}$ where the Fermi energy lies. A second-order perturbation reveals that an electron can jump between wires via a hopping process $\sim {\gamma^2}/{V_1}$ or $\sim {\gamma^2}/{V_2}$, which leads to an effective Hamiltonian of the form 

\begin{widetext}
\begin{dmath}
H_{\text{eff}} \sim  \sum_{k} \Bigg[ \Big( \sum_{\alpha=A,B}
\sum_{n=0}^{N-1}   E(k) c^\dagger_{k,n\alpha} c_{k,n\alpha} \Big) - \sum_{n=0}^{N-1} \frac{2\gamma^2}{V_1L} (c^\dagger_{k,nA} c_{k,n-1B} + \text{h.c.})  - \sum_{n=0}^{N-1} \frac{2\gamma^2}{V_2L} (c^\dagger_{k,nA} c_{k,nB} + \text{h.c.}) \Bigg]
\end{dmath}

Fourier transforming in the unit cell index $n$, one obtains:
\begin{align}
H &=   \sum_{k} \sum_q 
\begin{pmatrix}
c^\dagger_{k,qA} & 
c^\dagger_{k,qB} 
\end{pmatrix}   
\begin{pmatrix}
E(k) & \frac{2\gamma^2}{V_2 L} + \frac{2\gamma^2}{V_1 L} e^{-i q L}  \\
\frac{2\gamma^2}{V_2 L}  + \frac{2\gamma^2}{V_1 L} e^{i q L} & E(k)
\end{pmatrix}
\begin{pmatrix}
c_{k,qA}\\
c_{k,qB} 
\end{pmatrix}
\end{align}
\end{widetext}
This shows that each miniband effectively behaves like an SSH model with effective BDI symmetry, even though the microscopic Hamiltonian lies in the AI symmetry class. 

\section{Effective Hamiltonian near the gap-closing point from the scattering matrix}
\label{app:scateff}
In this section, we derive the low-energy effective Hamiltonian near the gap-closing point by starting from the network model's scattering matrix. This a complementary approach to that of \cref{sec:pertth}.

From Eq.~\eqref{eq:eigen}, the eigenvalue equation can be written as
\begin{align}
\mathcal Q \mathbf{b} = e^{-2 i \epsilon} \mathbf{b},
\end{align}
where \(\mathcal Q = (S_{\mathrm{unit}} M)^2\), which can be thought of as a two-step evolution operator.
The matrix $\mathcal Q$ is block diagonal and can be expressed as
\begin{align}
\mathcal Q = e^{i(\phi_1 + \phi_2)}
\begin{pmatrix}
\mathscr{N} & 0 \\
0 & \mathscr{M}
\end{pmatrix}.
\end{align}

The explicit forms of $\mathscr{N}$ and $\mathscr{M}$ are:
\begin{widetext} \begin{align} \mathscr N = \begin{pmatrix} \cos\theta_1 \cos\theta_2 - e^{2 i L q}\sin\theta_1 \sin\theta_2 & i\left( \cos\theta_2 \sin\theta_1 + e^{-2 i L q}\cos\theta_1 \sin\theta_2 \right) \\[6pt] i\left( \cos\theta_2 \sin\theta_1 + e^{2 i L q}\cos\theta_1 \sin\theta_2 \right) & \cos\theta_1 \cos\theta_2 - e^{-2 i L q}\sin\theta_1 \sin\theta_2 \end{pmatrix} \\ \mathscr M = \begin{pmatrix} \cos\theta_1 \cos\theta_2 - e^{2 i L q}\sin\theta_1 \sin\theta_2 & i\left( e^{-2 i L q}\cos\theta_2 \sin\theta_1 + \cos\theta_1 \sin\theta_2 \right) \\[6pt] i\left( e^{2 i L q}\cos\theta_2 \sin\theta_1 + \cos\theta_1 \sin\theta_2 \right) & \cos\theta_1 \cos\theta_2 - e^{-2 i L q}\sin\theta_1 \sin\theta_2 \end{pmatrix} \end{align} \end{widetext}

Since both blocks lead to identical low-energy physics, we focus on $\mathscr{N}$.
We expand around the gap-closing point defined by
\(\theta_2 = \theta_1 + \delta\),
with $\delta$ infinitesimal, and expand momenta near
\( q = \frac{\pi}{2L} + k, \)
with small $k$. To leading order in $\delta$ and $k$, the matrix $\mathscr{N}$ takes the form
\begin{align}
\mathscr{N} \approx \mathbb{I}_2 - i H_{\mathrm{eff}},
\end{align}
which defines the effective Hamiltonian
\begin{align}
H_{\mathrm{eff}} =
\begin{pmatrix}
2kL \sin\theta_1 \,\eta
&
-\delta + 2 i kL \cos\theta_1 \,\eta
\\[6pt]
-\delta - 2 i kL \cos\theta_1 \,\eta
&
-2kL \sin\theta_1 \,\eta
\end{pmatrix},
\end{align}
where \(\eta = \delta \cos\theta_1 + \sin\theta_1\). To bring the Hamiltonian to a more transparent form, we perform a unitary rotation
\begin{align}
\mathcal{U} = \exp({- i \frac{\pi}{4} \sigma_y}), 
\qquad
H'_{\mathrm{eff}} = \mathcal{U} H_{\mathrm{eff}} \mathcal{U}^\dagger.
\end{align}
This yields
\begin{align}
H'_{\mathrm{eff}} =
\begin{pmatrix}
\delta
&
2 i kL e^{-i \theta_1} \,\eta
\\[6pt]
-2 i kL e^{i \theta_1} \,\eta
&
-\delta
\end{pmatrix}.
\end{align}
We note that at $k=0$, the quasi-energy spectrum is gapped $\Delta \epsilon = 2\delta$.
In real space, a zero-energy state decays as
\(
\psi(x) \sim e^{-\delta  x / L },
\).
Then the localization length is given by
\(
\xi \sim {1}/{|\theta_1 - \theta_2|}.
\)
Thus, the gap closing at $\theta_1 = \theta_2$ is accompanied by a diverging localization length.

\begin{figure}[htpb]
    \centering
    \includegraphics[width=0.9\linewidth]{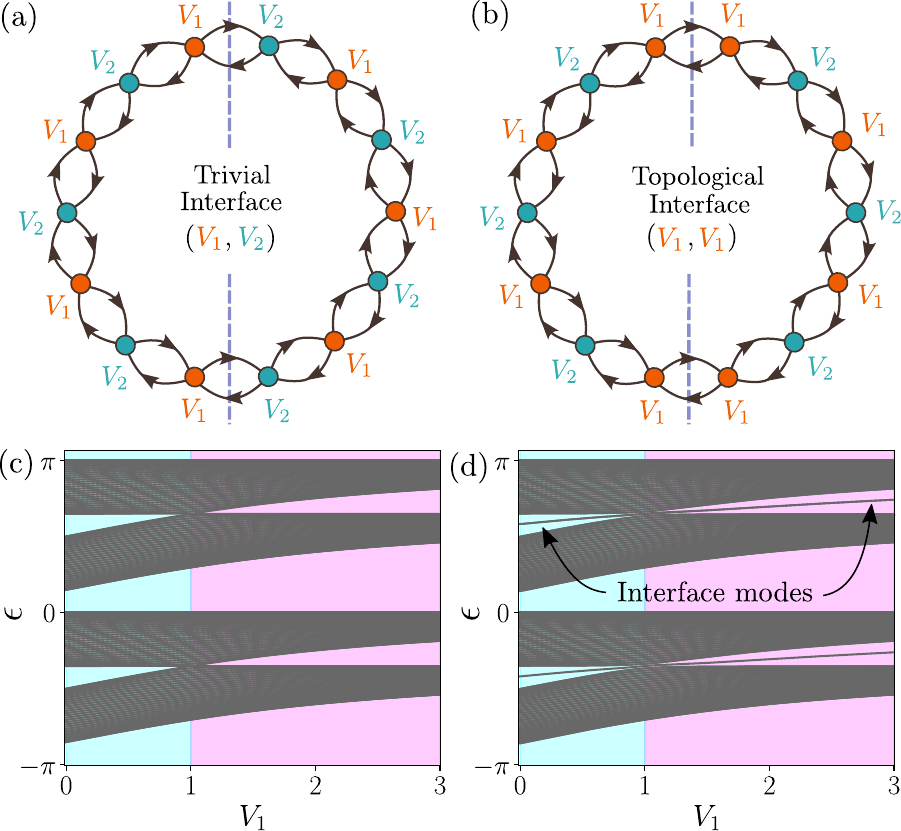}
    \caption{\textbf{Network interface configurations.} (a) Trivial interface configuration with alternating potentials $(V_1, V_2)$. (b) Topological interface configuration with identical potentials $(V_1, V_1)$ across the junction. (c)-(d) Quasi energy spectrum $\epsilon$ as a function of $V_1$ with $V_2=1$ and $N=60$ unit cells. (c) Spectrum corresponding to (a), showing no interface modes in the gap. (d) Spectrum corresponding to (b), exhibiting a localized interface mode within the gap. 
    }
    \label{fig:Interface}
\end{figure}

\section{Topological interface modes}
\label{app:interface}

In the main text (\cref{sec:mapping}), boundary potentials $V_L$ and $V_R$ were introduced to terminate the network. These parameters only implement the boundary conditions and do not affect the bulk topology. To make this explicit, we construct an interface geometry by joining two different configurations of the network without introducing any boundary potentials $(V_L, V_R)$ as shown in \cref{fig:Interface}(a)-(b). When the alternating structure $(V_1, V_2)$ is preserved across the interface \cref{fig:Interface}(a), the system remains topologically trivial and no sub-gap (interface) modes appear in the corresponding spectrum shown in \cref{fig:Interface}(c). In contrast, when identical potentials $(V_1, V_1)$ meet at the interface \cref{fig:Interface}(b), a localized interface mode appears within the gap, as shown in  \cref{fig:Interface}(d). Note that here $V_2=1$, implying that the impurity strength is kept at the order of hopping scale, so we are still within the metallic bandwidth. We have checked that this physics continues even when impurity strengths are much smaller than the $t$ scale since the topological transition of the network is decided by just relative impurity strengths (see \eqn{eq:rtfortb}).  This construction demonstrates that the interface mode arises from the mismatch of the bulk configurations and is independent of the boundary potentials $V_L$ and $V_R$.

\section{Reflection amplitudes for terminal impurity}
\label{app:rtends}

We derive the reflection $r_{R}$ and transmission amplitudes $t_{R}$ for an on-site potential $V_{R}$ located at the right end of a tight-binding chain. To avoid the additional propagation phase, we place the impurity at site $0$, and consider the semi-infinite tight-binding chain extending over sites $n \leq 0$.
The Hamiltonian is 
\beq
H= \sum_{n=-\infty}^{-1} -\gamma \Big( c^\dagger_{n} c_{n+1} + \text{h.c.} \Big) + V_{R} c^\dagger_{0} c_{0}
\eeq
In the bulk, for $n\leq-1$, the Schrodinger equation solution is
\beq
E\psi_{n}= -\gamma (\psi_{n+1}+ \psi_{n-1})
\eeq
with a scattering solution of incoming and reflected waves
\beq
\psi_{n}= e^{i k n} + r_{R} e^{-i k n}
\label{eq:bulkschrosol}
\eeq
and energy dispersion $E(k)=-2\gamma \cos k.$
At the impurity site, the Schrodinger equation becomes 
\beq
E\psi_{0}= -\gamma \psi_{-1} + V_{R}\psi_{0}
\eeq
Using \cref{eq:bulkschrosol}, $\psi_0=1+r_{R}$ and $\psi_{-1}= e^{-i k} + r_{R} e^{i k}$ with bulk dispersion $E(k)$, the reflection amplitude $r_{r}$ is given by
\beq
r_{R}= -\dfrac{V_{R} + \gamma e^{ik}}{V_{R} + \gamma e^{-ik}}.
\eeq
The reflection is unitary with $|r_{R}|=1$ and vanishing transmission $|t_{R}|=0$. So the boundary potential affects the phase of $r_{R}$, which is essential for reproducing the correct open-boundary spectrum with the network model description.

\bibliography{refs.bib}
\end{document}